\documentclass[12pt,a4paper]{article}
\pdfoutput=1

\usepackage[utf8]{inputenc}
\usepackage{color}
\usepackage{subcaption}
\usepackage[a4paper]{geometry}
\usepackage{graphicx}
\usepackage{amsmath}
\usepackage{amssymb}
\usepackage{float}
\usepackage{listings}
\usepackage{tikz}
\pdfoutput=1
\usepackage{epsfig,epsf}
 \usepackage{setspace}
 \usepackage{amsmath}
 \usepackage{array}
  \usepackage{tabu}
  \usepackage{graphicx}
\usepackage{subcaption}
\def\bea{\begin{eqnarray}}
\def\eea{\end{eqnarray}}

\title{\bf Open Spin Chains and Complexity \\ in the High Energy Limit}
\author{Grigorios Chachamis \&  Agust{\' \i}n Sabio Vera\\ \\
{\small Instituto de F{\' \i}sica Te{\' o}rica UAM/CSIC, Nicol{\'a}s Cabrera 15}\\ 
{\small \& Universidad Aut{\' o}noma de Madrid, E-28049 Madrid, Spain.}\\
}

\begin{document}

\maketitle 

\abstract
In the high energy limit of scattering amplitudes in Quantum Chromodynamics and supersymmetric theories the dominant Feynman diagrams
are characterized by a hidden integrability. A well-known example is that of Odderon exchange, which can be described as a bound state of three reggeized gluons and corresponds to a closed spin chain with periodic boundary conditions. 
In the $N=4$ supersymmetric Yang-Mills theory a similar spin chain arises in the multi-Regge asymptotics of the eight-point amplitude in the planar limit. We investigate the associated open spin chain in transverse momentum and rapidity variables solving the corresponding effective Feynman diagrams. We introduce the concept of complexity in the high energy effective field theory and study its emerging scaling laws. 
\begin{flushright}
{\sl Dedicated to the memory of Lev Lipatov}
\end{flushright}

\section{Introduction}

\begin{figure}[H] 
\vspace{-.51cm}
    \centering
    \includegraphics[width=0.35\linewidth]{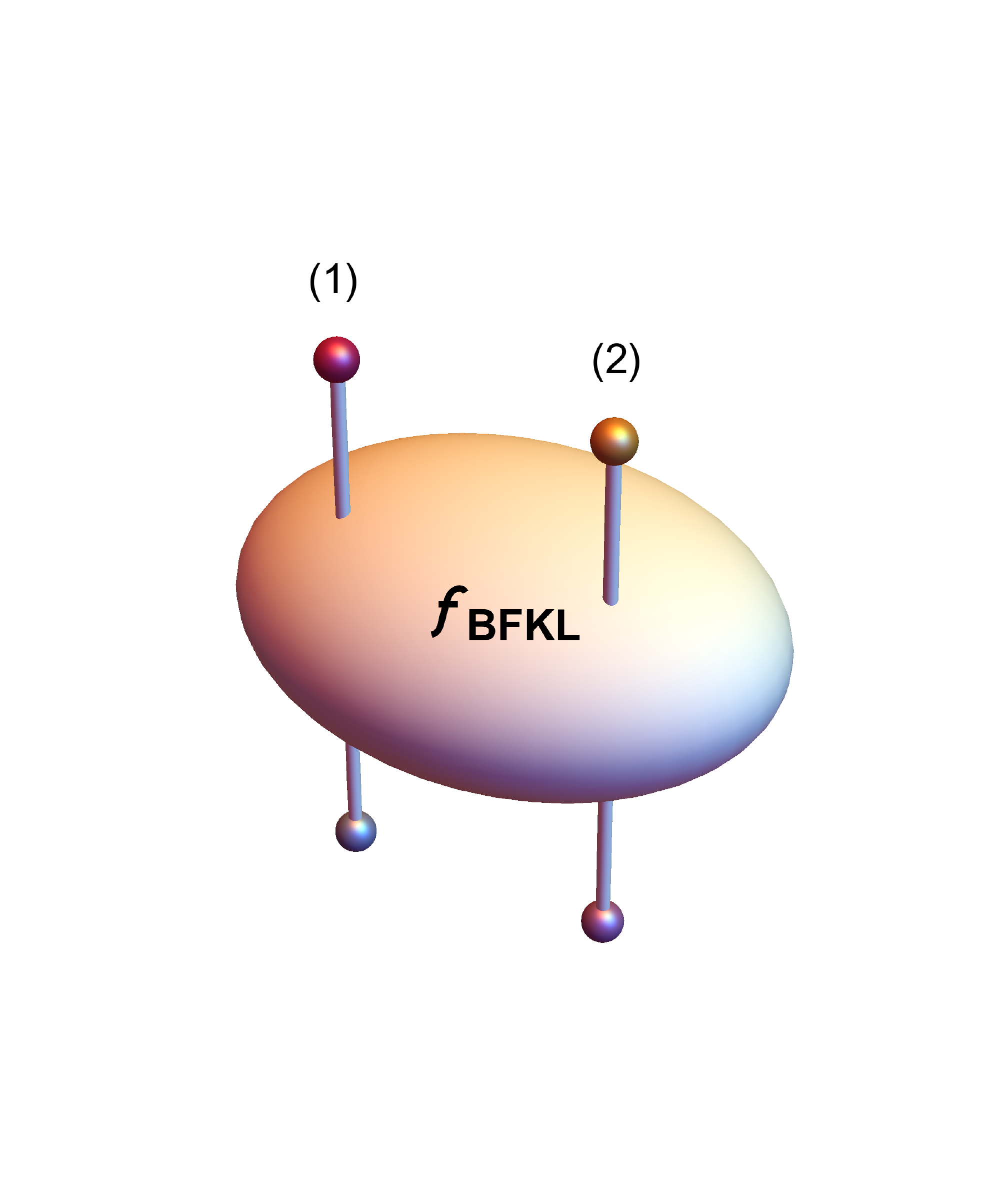} \includegraphics[width=0.35\linewidth]{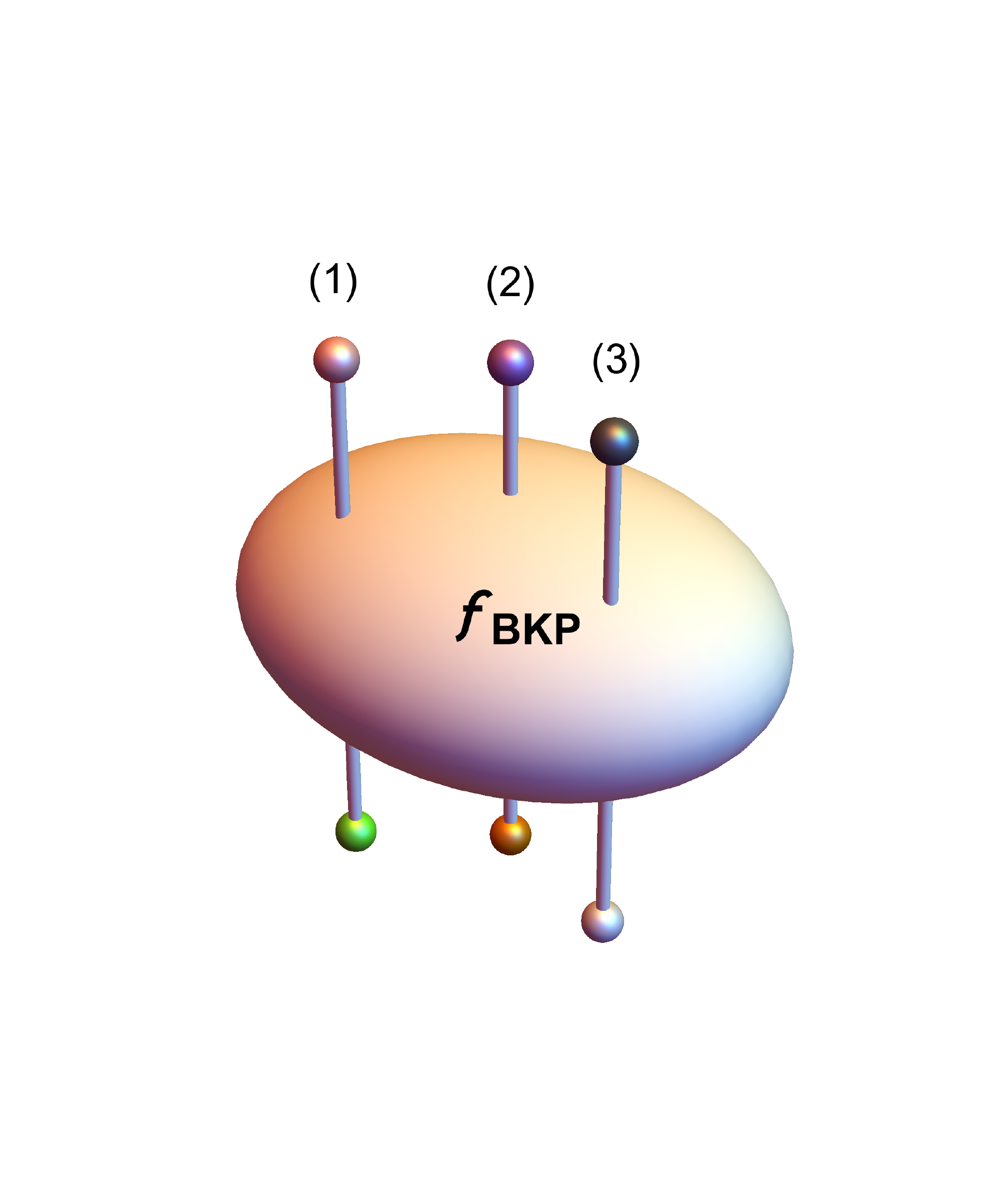} 
    \vspace{-1.3cm}
    \caption{Green functions describing the bound state of two (BFKL Pomeron) or three (BKP Odderon) reggeized-gluon bound states. } 
    \label{comb} 
\end{figure}

The high energy regime in Quantum Chromodynamics (QCD) has attracted a lot of attention for many years due to the rich structure of the theory in this limit. Novel effective degrees of freedom, reggeized quarks and gluons, arise when evaluating scattering amplitudes in the multi-Regge kinematical region (MRK). They can form bound states with different quantum numbers among which the best known are the Pomeron and the Odderon~\cite{Odderon}. The (perturbative) hard Pomeron (even under charge  and parity  conjugation) is best described  within the Balitsky-Fadin-Kuraev-Lipatov (BFKL) formalism~\cite{BFKL}. It corresponds to the dominant contribution to cross sections at very large energies while the Odderon (odd under charge and parity conjugation) is only important in a low energy window.  In perturbation theory they correspond, respectively, to a bound state of two and three reggeized gluons (see Fig.~\ref{comb}). A large amount of effort has been expended to study the properties of the Pomeron while for the Odderon the understanding, both theoretically and phenomenologically, is more limited~\cite{N.Cartiglia:2015gve}. 
The contribution to a cross section due to  Odderon exchange can be calculated by solving an integral equation describing the $t$-channel exchange of three off-shell reggeized gluons, the so-called Bartels-Kwiecinski-Praszalowicz  (BKP) equation~\cite{BKP}. Since, in MRK, this equation is invariant under two-dimensional conformal transformations in  coordinate representation, one can apply techniques developed in conformal field theories and integrable systems~\cite{Integrability,Recent} to find the solution to the equation. It is remarkable the existence of a mapping to a closed spin chain (CSC) when working in impact parameter representation~\cite{Integrability}. However, due to the different possible choices of normalization conditions for the Odderon wave function, it is difficult to know which analytic solution is relevant for a particular scattering process. 

In a recent work~\cite{Chachamis:2016ejm}, we solved the Odderon problem using an orthogonal method to those applied so far in the literature on this subject. Working in transverse momentum and rapidity space we solved exactly the equation governing the Odderon Green function. Our solution was expressed as a set of nested integrations which can be evaluated using numerical Monte Carlo techniques. 

The solution to the BKP equation for three Reggeons 
corresponds to a six-point amplitude with off-shells gluons $f_\omega \left({\bf p}_1,{\bf p}_2,{\bf p}_3;{\bf p}_4,{\bf p}_5,{\bf p}_6\right)$ (which we write as $f_\omega \left({\bf p}_1,{\bf p}_2,{\bf p}_3 \right)$). ${\bf p}_i$ are Euclidean two-dimensional transverse vectors where ${\bf p}_{i=1,2,3}$ have rapidity $Y$ and ${\bf p}_{i=4,5,6}$ a rapidity 0.  $\omega$ is a complex variable, Mellin-conjugate of $Y$. The reggeized gluon propagators are infrared divergent and need a regulator with mass-dimension $\lambda$.  With this regularization, the gluon Regge trajectory at leading order can be written in the form (with ${\bar \alpha}_s = \alpha_s N_c / \pi$)
\begin{eqnarray}
\omega ({\bf p}) &=& - \frac{{\bar \alpha}_s}{2} \ln{\frac{{\bf p}^2}{\lambda^2}}.
\end{eqnarray}
When the square of a Lipatov's emission vertex,  
\begin{eqnarray}
\xi \left({\bf p}_i,{\bf p}_j,{\bf p}_k,{\bf k}\right) ~= \frac{{\bar \alpha}_s}{4} \frac{ \theta({\bf k}^2 - \lambda^2)}{\pi {\bf k}^2}
 \Bigg(1+\frac{({\bf p}_i+{\bf k})^2 {\bf p}_j^2-({\bf p}_i+{\bf p}_j)^2 {\bf k}^2}{{\bf p}_i^2 ({\bf k}-{\bf p}_2)^2}\Bigg),
 \label{xifunction}
\end{eqnarray}
is integrated over the transverse momentum of the gluon with momentum ${\bf k}$ it generates a further $\lambda$ dependence which cancels that of the gluon trajectory.  The function $\xi$ in Eq.~(\ref{xifunction}) couples two reggeized gluons with momenta ${\bf p}_i$ and ${\bf p}_j$ via a $s$-channel normal gluon with transverse momentum ${\bf k}$, leaving the third reggeized gluon, with momentum ${\bf p}_k$, as a spectator. This generates pairwise nearest neighbor interactions in the corresponding CSC. 

The BKP equation in the Odderon case combines all these elements making use of a ternary kernel:
\begin{eqnarray}
\left(\omega - \omega({\bf p}_1) - \omega({\bf p}_2) - \omega({\bf p}_3) \right) 
 f_\omega \left({\bf p}_1,{\bf p}_2,{\bf p}_3\right) &=& \nonumber\\
&& \hspace{-6cm} \delta^{(2)}  \left({\bf p}_1-{\bf p}_4 \right) \delta^{(2)}  \left({\bf p}_2-{\bf p}_5 \right) 
 \delta^{(2)}  \left({\bf p}_3-{\bf p}_6 \right)  \nonumber\\
 &&\hspace{-6cm}+  \int d^2 {\bf k} \, \xi \left({\bf p}_1,{\bf p}_2,{\bf p}_3,{\bf k}\right)
 f_\omega \left({\bf p}_1+{\bf k},{\bf p}_2-{\bf k},{\bf p}_3\right) \nonumber\\
 &&\hspace{-6cm}+  \int d^2 {\bf k} \,  \xi \left({\bf p}_2,{\bf p}_3,{\bf p}_1,{\bf k}\right)
  f_\omega \left({\bf p}_1,{\bf p}_2+{\bf k},{\bf p}_3-{\bf k} \right) \nonumber\\
 &&\hspace{-6cm}+  \int d^2 {\bf k}  \, \xi \left({\bf p}_1,{\bf p}_3,{\bf p}_2,{\bf k}\right)f_\omega \left({\bf p}_1+{\bf k},{\bf p}_2,{\bf p}_3-{\bf k}\right).
\label{BKPeq} 
\end{eqnarray}
The initial condition for evolution in rapidity corresponds to three reggeized-gluon propagators normalized in the form of  three two-dimensional Dirac delta functions $\delta^{(2)}  \left({\bf p}_1-{\bf p}_4 \right) \delta^{(2)}  \left({\bf p}_2-{\bf p}_5 \right)  \delta^{(2)}  \left({\bf p}_3-{\bf p}_6 \right)$.

In Ref.~\cite{Chachamis:2016ejm} we showed how to iterate the BKP ternary kernel acting on the initial condition until we reach convergence for a particular value of the relevant expansion parameter  ${\bar \alpha}_s Y$.  The gluon Green function  grows with $Y$ for small values of this variable to then rapidly decrease at higher $Y$ (keeping ${\bar \alpha}_s$ constant). Our solution is compatible with previous approaches where the Odderon intercept has been argued to be of ${\cal O} (1)$~\cite{Bartels:1999yt} (similar results have been found within the dipole formalism~\cite{Kovchegov:2003dm}).  

In the following section we will explain in some detail our procedure to solve an equation similar to Eq.~(\ref{BKPeq}) which has a representation as an integrable open spin chain (OSC) as shown by Lev N. Lipatov in 
Ref.~\cite{Lipatov:2009nt}. In a nutshell, we first iterate it in the $\omega$ space to then transform the result to get back to a representation with only transverse momenta and rapidity. 

\section{The equation for the open spin chain}

When evaluating the eight-gluon amplitude in the $N=4$ supersymmetric theory in MRK  and in certain physical regions we encounter a contribution with three Reggeized gluons exchanged in the $t$-channel. This is the most complicated contribution to the amplitude stemming from the so-called Mandelstam cuts in the associated partial wave~\cite{Bartels:2008ce}. Since the amplitude carries the quantum numbers of a gluon in that channel, this implies that the contributing effective Feynman diagrams are planar and the gluons with 
momentum ${\bf p}_1$  and ${\bf p}_3$ cannot be directly connected by the function $\xi$. This is the reason why we now have an OSC. The corresponding BKP-like integral equation then reads
\begin{eqnarray}
\left(\omega - \omega({\bf p}_1) - \omega({\bf p}_2) - \omega({\bf p}_3) \right) 
 f_\omega \left({\bf p}_1,{\bf p}_2,{\bf p}_3\right) &=& \nonumber\\
&& \hspace{-6cm} \delta^{(2)}  \left({\bf p}_1-{\bf p}_4 \right) \delta^{(2)}  \left({\bf p}_2-{\bf p}_5 \right) 
 \delta^{(2)}  \left({\bf p}_3-{\bf p}_6 \right)  \nonumber\\
 &&\hspace{-6cm}+  \int d^2 {\bf k} \, \xi \left({\bf p}_1,{\bf p}_2,{\bf p}_3,{\bf k}\right)
 f_\omega \left({\bf p}_1+{\bf k},{\bf p}_2-{\bf k},{\bf p}_3\right) \nonumber\\
 &&\hspace{-6cm}+  \int d^2 {\bf k} \,  \xi \left({\bf p}_2,{\bf p}_3,{\bf p}_1,{\bf k}\right)
  f_\omega \left({\bf p}_1,{\bf p}_2+{\bf k},{\bf p}_3-{\bf k} \right).
\label{OpenBKP} 
\end{eqnarray}
Following closely our approach for the CSC, we can solve this equation by iteration. With this in mind, it is convenient to use the operator notation
\begin{eqnarray}
{\cal O} ({\bf k}) \otimes f \left({\bf p}_1,{\bf p}_2,{\bf p}_3\right) & \equiv& \xi \left({\bf p}_1,{\bf p}_2,{\bf p}_3,{\bf k}\right)
 f \left({\bf p}_1+{\bf k},{\bf p}_2-{\bf k},{\bf p}_3\right) \nonumber\\
 &+& \xi \left({\bf p}_2,{\bf p}_3,{\bf p}_1,{\bf k}\right)
  f \left({\bf p}_1,{\bf p}_2+{\bf k},{\bf p}_3-{\bf k} \right),
 \label{eq:bkpkernel} 
\end{eqnarray}
to write the equation in the form
\begin{eqnarray}
\left(\omega - \omega({\bf p}_1) - \omega({\bf p}_2) - \omega({\bf p}_3) \right) 
 f_\omega \left({\bf p}_1,{\bf p}_2,{\bf p}_3\right) &=& \nonumber\\
&& \hspace{-8.cm} \delta^{(2)}  \left({\bf p}_1-{\bf p}_4 \right) \delta^{(2)}  \left({\bf p}_2-{\bf p}_5 \right) 
 \delta^{(2)}  \left({\bf p}_3-{\bf p}_6 \right) \nonumber\\
&& \hspace{-7.cm} +\int d^2 {\bf k} \,  {\cal O} ({\bf k}) \otimes f_\omega \left({\bf p}_1,{\bf p}_2,{\bf p}_3\right).
\end{eqnarray}
Iterating this expression we obtain
\begin{eqnarray}
 f_\omega \left({\bf p}_1,{\bf p}_2,{\bf p}_3\right) &=&   \frac{ \left(1+\sum_{n=1}^\infty \prod_{i=1}^n \int d^2 {\bf k}_i {\cal O} ({\bf k}_i) \otimes \right)  }{\left(\omega - \omega({\bf p}_1) - \omega({\bf p}_2) - \omega({\bf p}_3) \right) } \nonumber\\
 &&\delta^{(2)}  \left({\bf p}_1-{\bf p}_4 \right) \delta^{(2)}  \left({\bf p}_2-{\bf p}_5 \right) \delta^{(2)}  \left({\bf p}_3-{\bf p}_6 \right).
\end{eqnarray}
We can now trade $\omega$  for a $Y$ dependence using
\begin{eqnarray}
f \left({\bf p}_1,{\bf p}_2,{\bf p}_3, Y\right) &=& \int_{a-i \infty}^{a + i \infty} \frac{d \omega}{2 \pi i}  e^{\omega Y} 
f_\omega \left({\bf p}_1,{\bf p}_2,{\bf p}_3\right)  
\end{eqnarray}
and write our solution in the form
\begin{eqnarray}
f_{\rm BKP}^{\rm adjoint} \left({\bf p}_1,{\bf p}_2,{\bf p}_3, Y\right)  &=&   \nonumber\\
&&\hspace{-4cm}e^{(\omega({\bf p}_1) + \omega({\bf p}_2) + \omega({\bf p}_3)) Y} \delta^{(2)}  \left({\bf p}_1-{\bf p}_4 \right) \delta^{(2)}  \left({\bf p}_2-{\bf p}_5 \right) \delta^{(2)}  \left({\bf p}_3-{\bf p}_6 \right)
\nonumber\\
 &&\hspace{-4cm} + \sum_{n=1}^\infty \Bigg\{\prod_{i=1}^n \int_0^{y_{i-1}} d y_i 
 \int d^2 {\bf k}_i e^{(\omega({\bf p}_1) + \omega({\bf p}_2) + \omega({\bf p}_3)) (y_{i-1}- y_i)} {\cal O} ({\bf k}_i) \otimes \Bigg\} \nonumber\\
 &&\hspace{-3.5cm}e^{(\omega({\bf p}_1) + \omega({\bf p}_2) + \omega({\bf p}_3)) y_n} 
 \delta^{(2)}  \left({\bf p}_1-{\bf p}_4 \right) \delta^{(2)}  \left({\bf p}_2-{\bf p}_5 \right) \delta^{(2)}  \left({\bf p}_3-{\bf p}_6 \right).
 \label{IterativeEqn}
\end{eqnarray}
It is important to comment now on the infrared finiteness of this expression. The original non-forward BFKL equation is projected on a color singlet representation and reads
\begin{eqnarray}
\left(\omega - \omega({\bf p}_1) - \omega({\bf p}_2) ) \right) 
 f_\omega \left({\bf p}_1,{\bf p}_2\right) &=&  \delta^{(2)}  \left({\bf p}_1-{\bf p}_3 \right)  \nonumber\\
 &&\hspace{-5cm}+ \, 2  \int d^2 {\bf k} \, \xi \left({\bf p}_1,{\bf p}_2,{\bf k}\right)
 f_\omega \left({\bf p}_1+{\bf k},{\bf p}_2-{\bf k}\right).
\label{BFKLsinglet} 
\end{eqnarray}
The solution to this equation is $\lambda$ independent in the limit $\lambda \to 0$ because each of the logarithmic divergencies generated upon integration over one of the gluons with momentum ${\bf k}$ is compensated by the same logarithmic $\lambda$ dependence in two of the gluon Regge trajectories. The original BKP equation (for any number of exchanged reggeons) projected on a singlet is IR finite for the same reason. For example, the divergence generated by 
\begin{eqnarray}
 \int d^2 {\bf k} \, \xi \left({\bf p}_1,{\bf p}_2,{\bf p}_3,{\bf k}\right)
 f_\omega \left({\bf p}_1+{\bf k},{\bf p}_2-{\bf k},{\bf p}_3\right) 
\end{eqnarray}
in Eq.~(\ref{BKPeq}) is cancelled against the one present in  $ \left(\omega({\bf p}_1) + \omega({\bf p}_2) \right)/2$. 

The situation is different when we project on the adjoint representation in the $t$-channel. Now the corresponding Green functions are IR divergent. Fortunately, the associated $\lambda$ dependence factorizes in a simple form. Let us go back to the BFKL equation in this new color representation, it now reads
\begin{eqnarray}
\left(\omega - \omega({\bf p}_1) - \omega({\bf p}_2) ) \right) 
 f_\omega \left({\bf p}_1,{\bf p}_2\right) &=&  \delta^{(2)}  \left({\bf p}_1-{\bf p}_3 \right)  \nonumber\\
 &&\hspace{-5cm}+  \int d^2 {\bf k} \, \xi \left({\bf p}_1,{\bf p}_2,{\bf k}\right)
 f_\omega \left({\bf p}_1+{\bf k},{\bf p}_2-{\bf k}\right).
\label{BFKLadjoint} 
\end{eqnarray}
We can see that there is a divergent $\lambda$ dependence generated by half of the trajectories contribution 
$ \left(\omega({\bf p}_1) + \omega({\bf p}_2) \right)/2$ not compensated by the integration of $\xi$. In Ref.~\cite{BFKLex1} we have shown that in this case the gluon Green function can be written as
\begin{eqnarray}
f_{\rm BFKL}^{\rm adjoint} \left({\bf p}_1,{\bf p}_2, {\rm Y}\right) &=& \left(\frac{\lambda^2}{\sqrt{{\bf p}_1^2 {\bf p}_2^2}}\right)^{\frac{\bar{\alpha}_s {\rm Y}}{2}} {\widehat f}_{\rm BFKL}^{\, \, {\rm adjoint}} \left({\bf p}_1,{\bf p}_2, {\rm Y}\right), 
\end{eqnarray}
where ${\widehat f}_{\rm BFKL}^{\, \, {\rm adjoint}} $ is finite when $\lambda \to 0$. As we mentioned before, the usual BKP equation for three reggeized gluons is finite while the open BKP-like equation we are studying lacks the link term $\xi$ between ${\bf p}_1$ and ${\bf p}_3$. This means that the combination of trajectories $ \left(\omega({\bf p}_1) + \omega({\bf p}_3) \right)/2$ generates divergencies which are not cancelled. Similarly to the adjoint BFKL case these divergencies factorize at the level of the Green function in the form
\begin{eqnarray}
f_{\rm BKP}^{\rm adjoint} \left({\bf p}_1,{\bf p}_2,{\bf p}_3, {\rm Y}\right) &=& \left(\frac{\lambda^2}{\sqrt{{\bf p}_1^2 {\bf p}_3^2}}\right)^{\frac{\bar{\alpha}_s {\rm Y}}{2}} {\widehat f}_{\rm BKP}^{\, \, {\rm adjoint}} \left({\bf p}_1,{\bf p}_2,{\bf p}_3, {\rm Y}\right).
\end{eqnarray}
It is the infrared finite function ${\widehat f}_{\rm BKP}$ the one we will study in depth in this work.  Let us now indicate that the $\lambda$ independence can be achieved order-by-order in a coupling expansion but this is not very useful for our purposes. We work instead with an effective field theory where each Feynman diagram is not infrared finite. Finiteness is achieved only after summing up all contributing effective Feynman diagrams to all those orders being numerically significant. We explained 
this method in detail in Ref.~\cite{BFKLex1}. We describe the main features emerging from our solution in the following section.

\section{Solution to the open spin chain}

In this section we evaluate the solution to the OSC and compare it to that in the closed case. 
It is instructive to discuss first a few key points regarding the type of diagrams one encounters in
the two cases. 
\begin{figure}[H] 
 \vspace{-.6cm}
    \centering
    \includegraphics[width=0.9\linewidth]{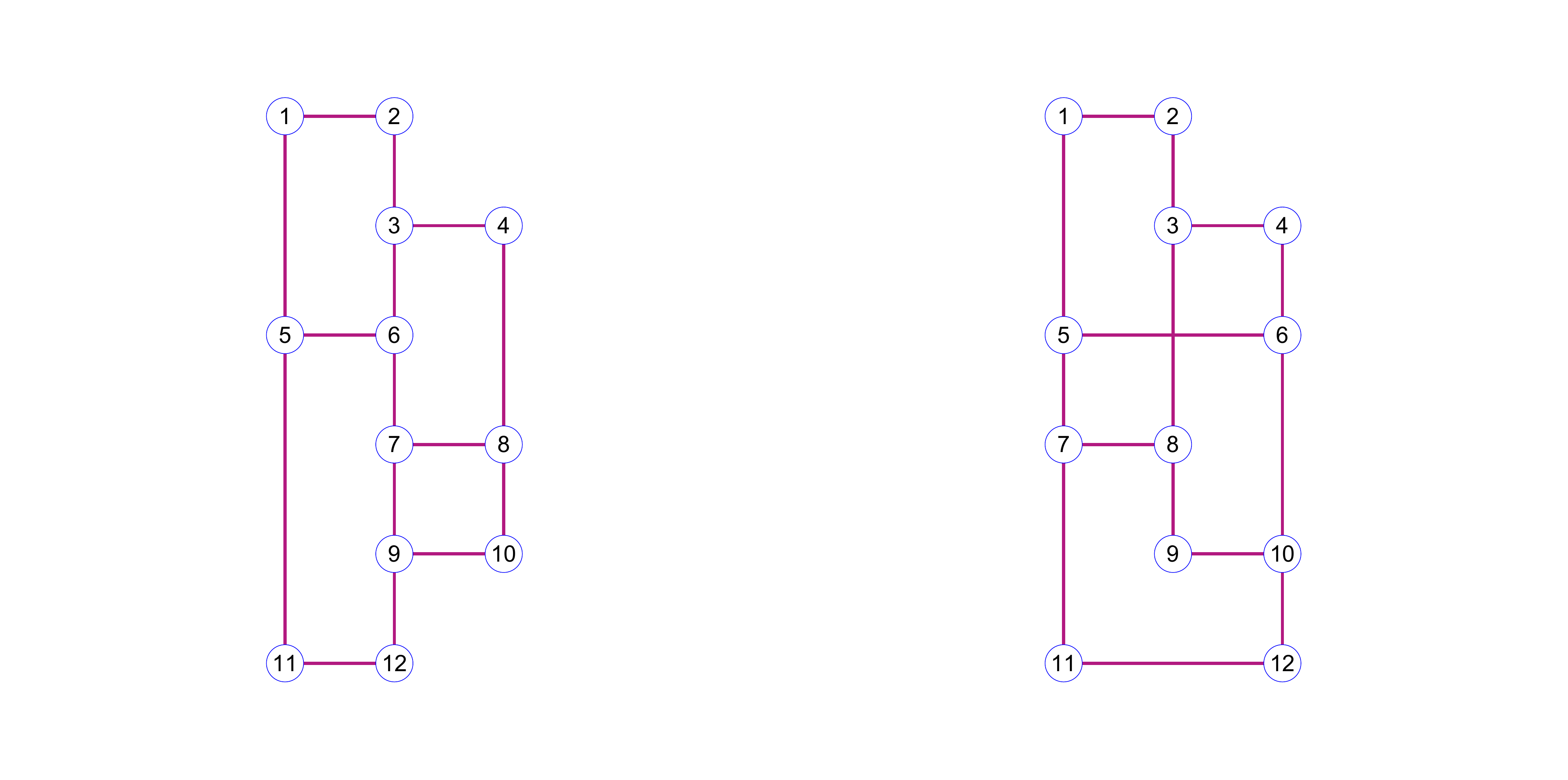} 
     \vspace{-.6cm}
    \caption{Two typical diagrams with six rungs. Open spin chain topology to the left 
    (labeled \{L, R, L, R, R, L\})
    and closed spin chain topology 
    to the right (labeled \{L, R, M, L, R, M\}).    } 
    \label{grafs2} 
\end{figure}

We will always have three Reggeons labeled (1), (2) and (3) vertically aligned and ordered
such that (1) is to the left and (3) to the right (see the plot at the right hand side of Fig.~\ref{comb}). In the CSC 
case each Reggeon is allowed to interact with any of the other two via gluon exchange. We will call these gluon exchanges {\it rungs} following the nomenclature associated to generalised ladder diagrams. We will label a rung between Reggeons (1) and (2) by L (left), a rung between Reggeons (2) and (3) by R (right) and a rung between Reggeons (1) and (3) by M.
As an example, in Fig.~\ref{grafs2} we can see two six-rung ladder diagrams, one for the OSC case (to the left) and one for the closed
case (to the right). In the following, without any loss of generality, we will assume that in the OSC case the Reggeons  that cannot interact directly via a gluon exchange are the (1) and (3). 

\begin{figure}[H] 
\vspace{-.4cm}
    \centering
    \includegraphics[width=0.25\linewidth]{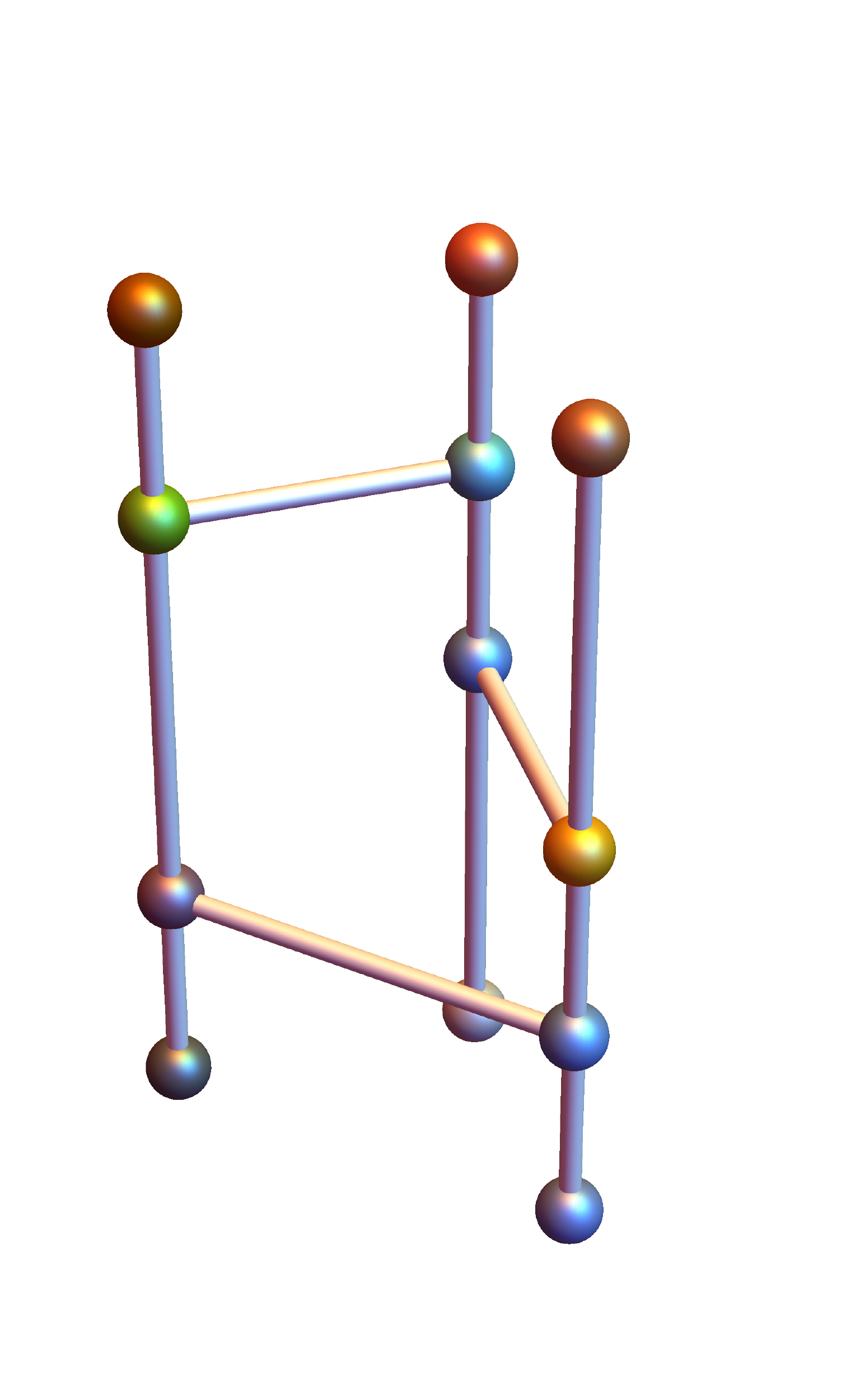}  \includegraphics[width=0.25\linewidth]{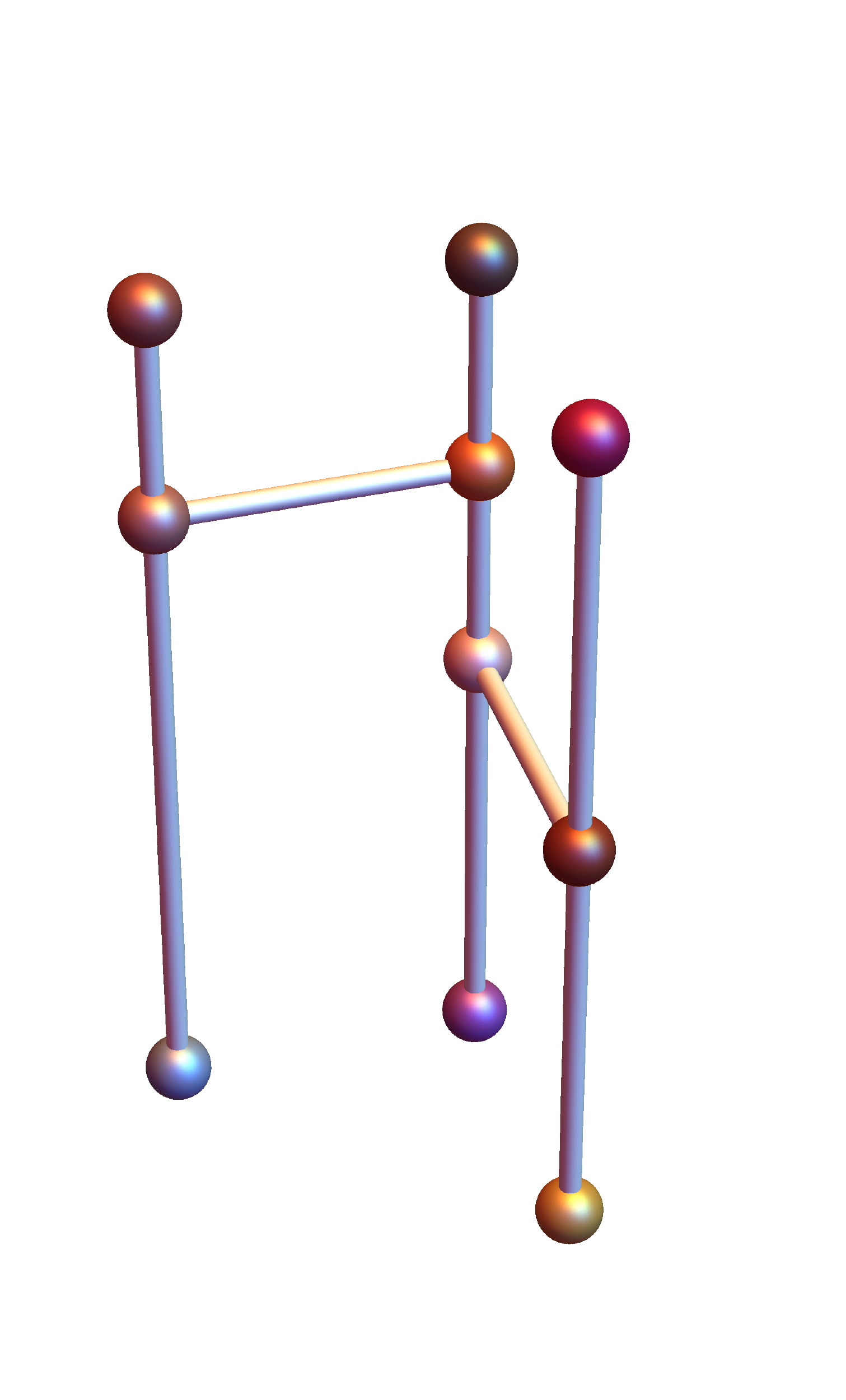} 
    \vspace{-.6cm}
    \caption{A closed spin chain diagram (left) has a cylindrical symmetry and cannot be embedded on a plane. Conversely,
    an open spin diagram (right), has no cylindrical symmetry and is planar.} 
    \label{grafs3} 
\end{figure}

From Fig.~\ref{grafs2}, it is evident that an ordered list of the labels that correspond to the rungs of a ladder diagram defines it unambiguously.  In our diagrams we only have two types of nodes: those with two attached legs and those with three. There are only six of the first type which correspond to three-point vertices where we have removed one propagator carrying one of the external momenta ${\bf p}_i$.  The {\it adjacency matrix} of a given diagram is the  square matrix with off-diagonal elements being the number of lines connecting the vertex $i$ with the vertex $j$. 
Its diagonal elements are zero.  Using a labelling system as described above  allows
the fast computation of the adjacency matrix and adjacency list for any ladder diagram opening the road for more detailed studies of the diagrammatic topologies that appear after each iteration of the kernel. As an example we now show the adjacency matrices associated to the diagrams in Fig.~\ref{grafs2}. For the CSC example,
\begin{eqnarray}
\left(
\begin{array}{cccccccccccc}
0&1&0&0&1&0&0&0&0&0&0&0\\
1&0&1&0&0&0&0&0&0&0&0&0\\
0&1&0&1&0&0&0&1&0&0&0&0\\
0&0&1&0&0&1&0&0&0&0&0&0\\
1&0&0&0&0&1&1&0&0&0&0&0\\
0&0&0&1&1&0&0&0&0&1&0&0\\
0&0&0&0&1&0&0&1&0&0&1&0\\
0&0&1&0&0&0&1&0&1&0&0&0\\
0&0&0&0&0&0&0&1&0&1&0&0\\
0&0&0&0&0&1&0&0&1&0&0&1\\
0&0&0&0&0&0&1&0&0&0&0&1\\
0&0&0&0&0&0&0&0&0&1&1&0\\
\end{array}
\right),
\end{eqnarray}
and for the open,
\begin{eqnarray}
\left(
\begin{array}{cccccccccccc}
0&1&0&0&1&0&0&0&0&0&0&0\\
1&0&1&0&0&0&0&0&0&0&0&0\\
0&1&0&1&0&1&0&0&0&0&0&0\\
0&0&1&0&0&0&0&1&0&0&0&0\\
1&0&0&0&0&1&0&0&0&0&1&0\\
0&0&1&0&1&0&1&0&0&0&0&0\\
0&0&0&0&0&1&0&1&1&0&0&0\\
0&0&0&1&0&0&1&0&0&1&0&0\\
0&0&0&0&0&0&1&0&0&1&0&1\\
0&0&0&0&0&0&0&1&1&0&0&0\\
0&0&0&0&1&0&0&0&0&0&0&1\\
0&0&0&0&0&0&0&0&1&0&1&0\\
\end{array}
\right).
\end{eqnarray}
We will make use of this representation of the Feynman diagrams in the following section. 

In Ref.~\cite{Chachamis:2016ejm}, we developed a Monte Carlo code to compute the Odderon Green function based on our
experience with {\tt BFKLex}~\cite{BFKLex1,BFKLex2}. In principle, a straightforward approach would be to 
modify as little as possible our existing code for the CSC case and run it for the OSC one. However,
we realized at an early stage of this work that we would need to develop a code tailored especially for the OSC case. This 
was mainly due to the fact that convergence now was not as fast as in the CSC case and therefore we had to 
optimise our code as much as possible. In the following we will describe some key issues regarding our numerical approach, 
for more details we refer 
the reader to Ref.~\cite{Chachamis:2016ejm}.

\begin{figure}[H] 
    \centering
    \includegraphics[width=0.75\linewidth]{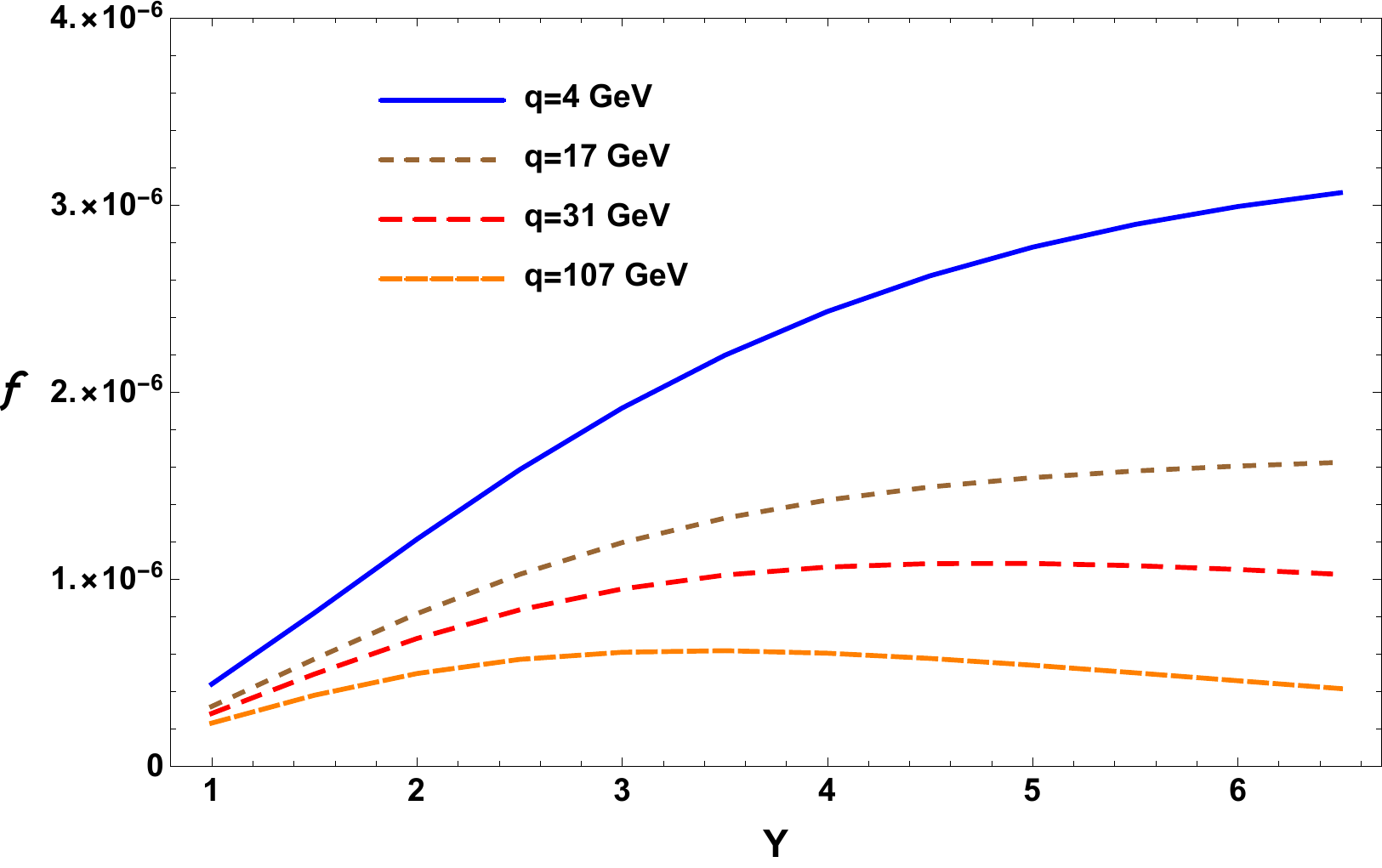} 
 \caption{Energy dependence of $f$ for $q = 4$ GeV (blue continuous line), $q = 31$ GeV (red dashed line), $q=17$ GeV (brown dashed line)  and $q=107$ GeV (green dashed line) for the open spin chain. }
  \label{en2} 
\end{figure}

We want to compute Eq.~(\ref{IterativeEqn})
in order to obtain  $f \left({\bf p}_1,{\bf p}_2,{\bf p}_3, Y\right)$ for some given momenta configuration
 ${\bf p}_1,{\bf p}_2,{\bf p}_3,{\bf p}_4,{\bf p}_5,{\bf p}_6$ and for varying rapidity $Y$. 
To be specific, we will consider the following values for the transverse 
momenta  (they are shown in polar coordinates,
the first entry stands for the modulus of the momentum and the second
one for the azimuthal angle):
 \begin{align} 
%& \,\, \, {\bf q} = (4, 0)                                                     &{\bf q} &= (31, 0); \nonumber \\
& {\bf p}_1 = (10, 0) ;                            \nonumber \\ %                  &{\bf p}_1  &= (10, 0); \nonumber \\
& {\bf p}_2  = (20, \pi);                            \nonumber \\ %                &{\bf p}_2  &= (20, \pi); \nonumber \\
& {\bf p}_3 = {\bf q}-{\bf p}_1-{\bf p}_2 ; \nonumber \\ % = (14,0)   &  {\bf p}_3 &= ({\bf q}-{\bf p}_1)-{\bf p}_2 = (41,0); \nonumber \\
& {\bf p}_4  = (20, 0);                                 \nonumber \\%              &{\bf p}_4  &= (20, 0) \nonumber \\
& {\bf p}_5 =   (25, \pi);                                \nonumber \\%            &{\bf p}_5  &= (25, \pi) \nonumber \\
& {\bf p}_6 = {\bf q}-{\bf p}_4 -{\bf p}_5.%=(9,0)        &  {\bf p}_6 &=({\bf q}-{\bf p}_4)-{\bf p}_5=(36,0).
\label{momenta}
 \end{align}
 The units of the moduli in the two-vectors above
 are expressed in GeV. 
 The  momentum transfer ${\bf q}$  satisfies
\begin{align}
{\bf p}_1 + {\bf p}_2 + {\bf p}_3 = {\bf q} = {\bf p}_3 + {\bf p}_4 + {\bf p}_6
\label{mom-equality}
\end{align}
and takes the following values: ${\bf q} = \{(4,0), (17,0), (31,0), (107,0)\}$.
 Finally, we  vary the rapidity $Y$ from 1 to 5.5 units and in one case up to 6.5 (Fig.~\ref{en2}).

As in the CSC case, non-zero contributions to $f \left({\bf p}_1,{\bf p}_2,{\bf p}_3, Y\right)$ will appear
only after we consider two rungs. Any given diagram with
$i$ rungs, once iterated, will generate two new diagrams each with $i+1$ rungs. This leads to a complete binary 
tree structure:
\begin{center}
\begin{tikzpicture}[every node/.style={circle,draw},level 1/.style={sibling distance=40mm},level 2/.style={sibling distance=20mm}
]
\node {0}
child { 
node {L} 
child { node {LL}}
child { node {LR}}
}
child { 
node {R} 
child {node {RL}}
child {node {RR}}
}
;
\label{tex-tree}
\end{tikzpicture}
\end{center}
Two of the momenta integrations
are trivial since there are three Dirac delta functions to be fulfilled.
However, the {\it junctions} (see Ref.~\cite{Chachamis:2016ejm}) 
in the OSC are only two:
${\mathcal{J}}_{\text{LR}}$,
${\mathcal{J}}_{\text{RL}}$.
Lastly, any diagram with no junction, in other words,  any diagram ${\mathcal{J}}_{\text{Q}}$
 with Q being a sequence of only L or only R is zero. This becomes clear since all ${\bf p}_i$ are chosen to be different from each other and hence none of the Dirac delta functions in the initial condition is  fulfilled.

\begin{figure}[H]
\centering
\includegraphics[width=0.75\linewidth]{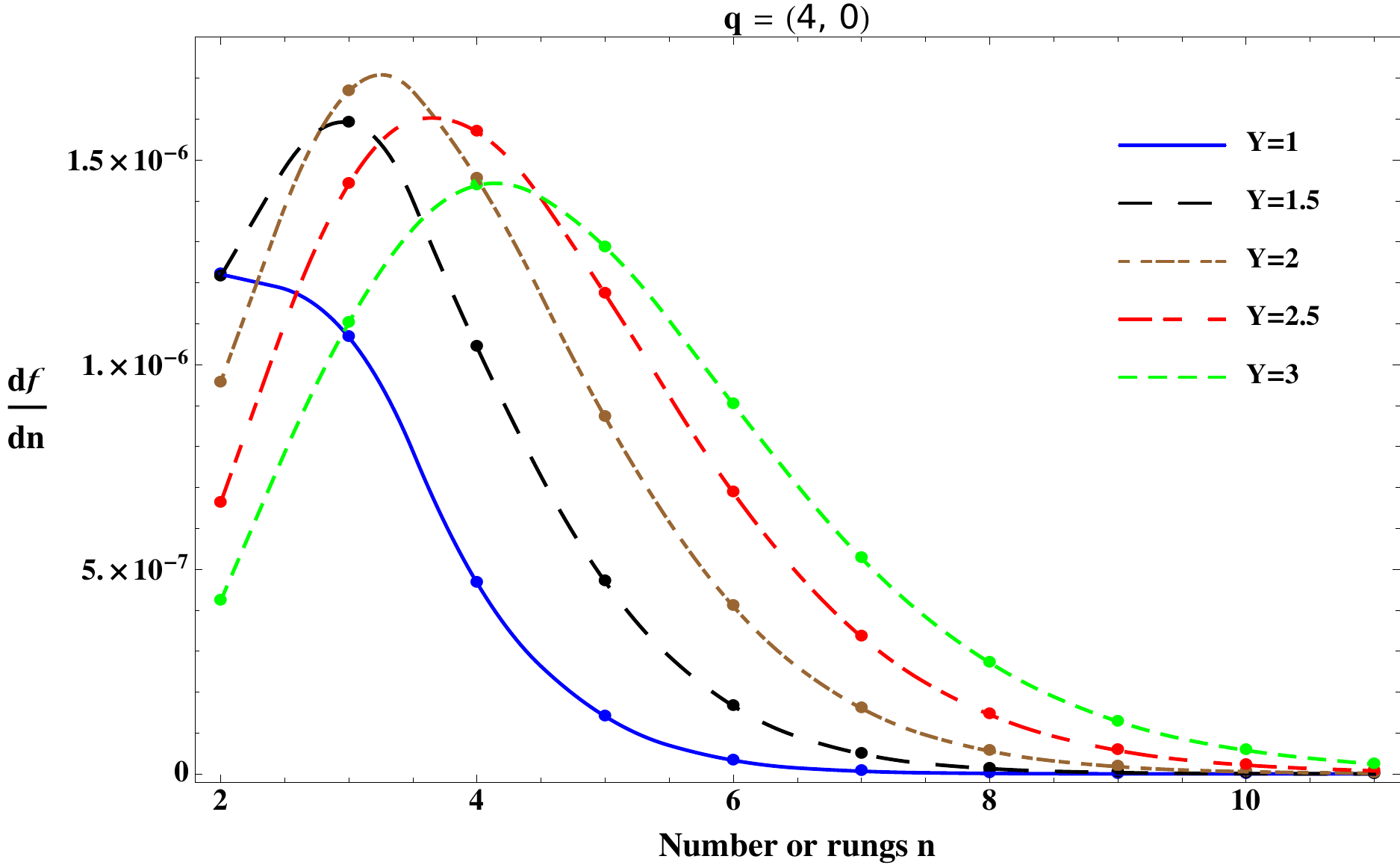} \\
\vspace{.5cm}
\includegraphics[width=0.75\linewidth]{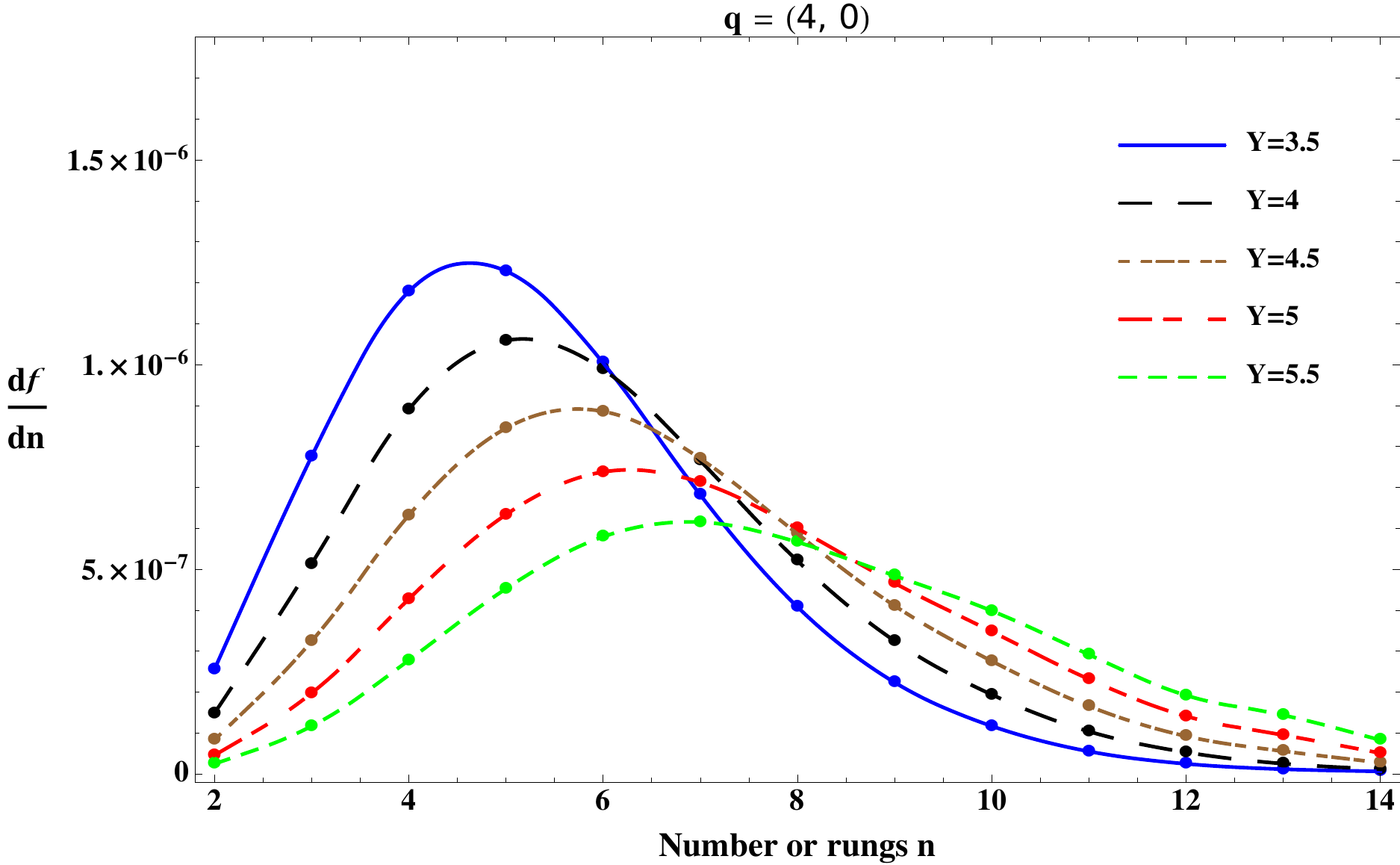} 
\caption{{\it Multiplicity} plot for $q = 4$ GeV in the closed spin chain.}  
    \label{q4-1} 
\end{figure}

\begin{figure}[H]
\centering
\includegraphics[width=0.75\linewidth]{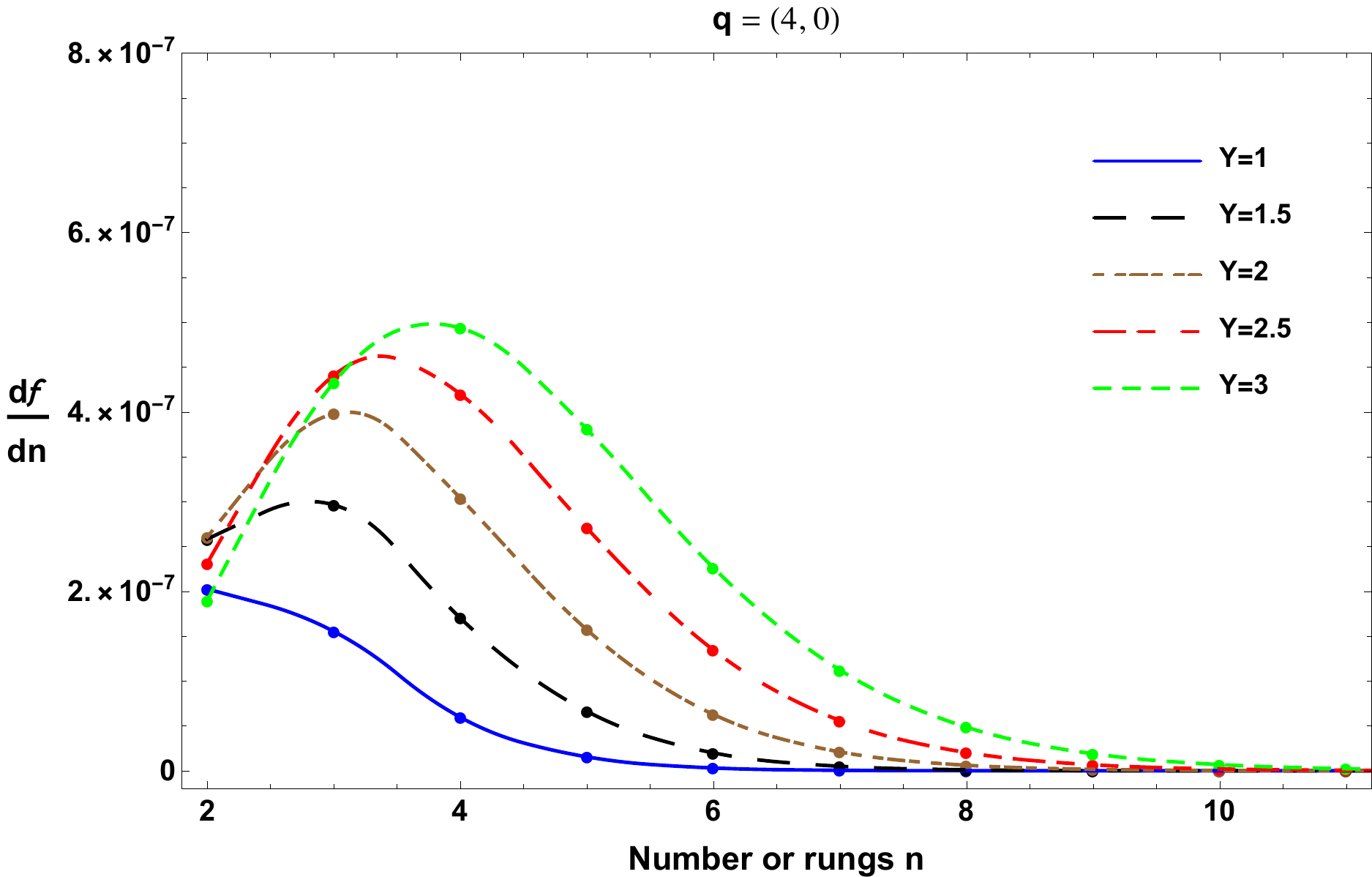}  \\
\vspace{.5cm}
    \includegraphics[width=0.75\linewidth]{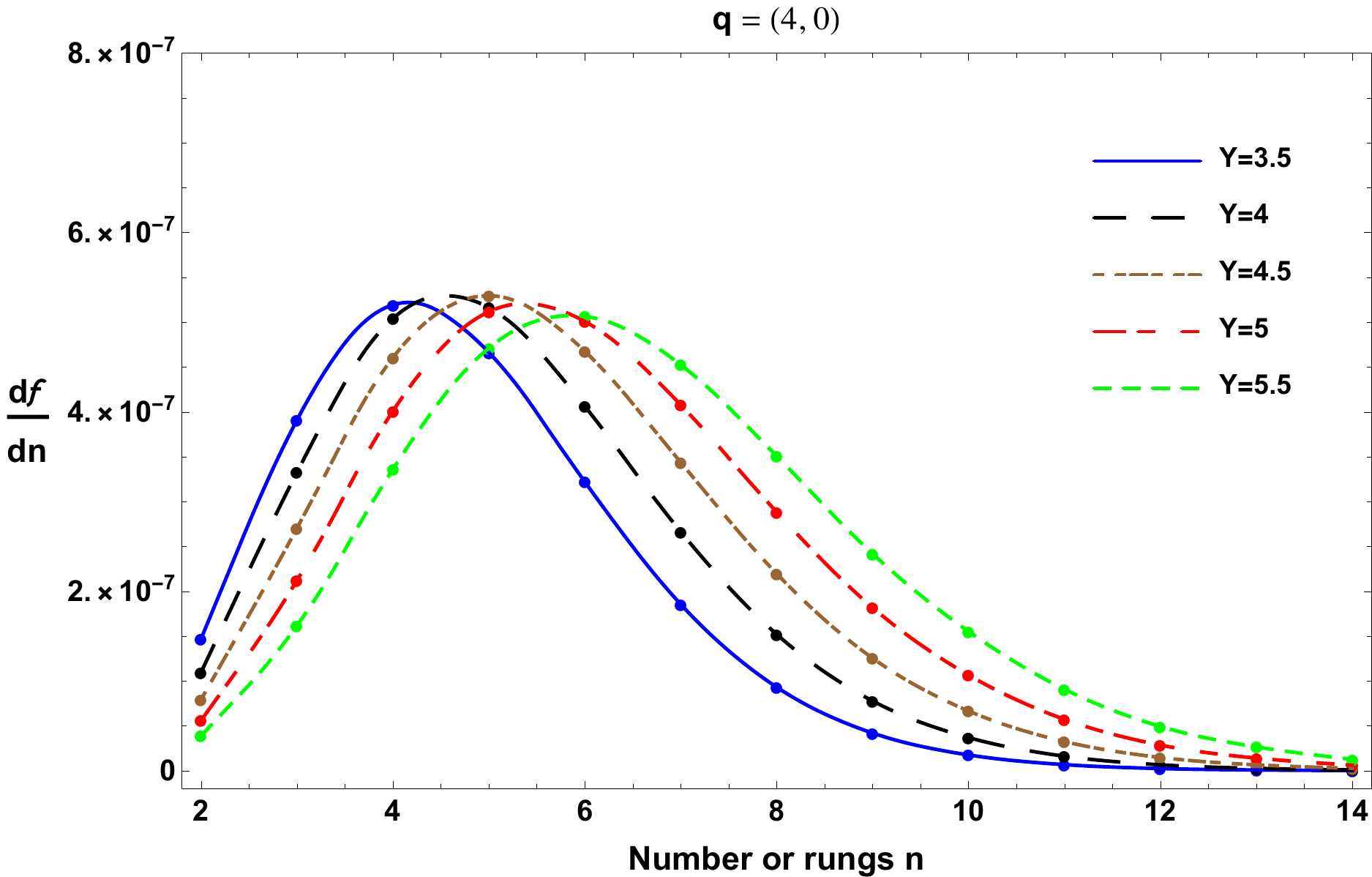}
\caption{{\it Multiplicity} plot for $q = 4$ GeV in the open spin chain.}  
    \label{q4-2} 
\end{figure}

\begin{figure}
\centering
\includegraphics[width=0.75\linewidth]{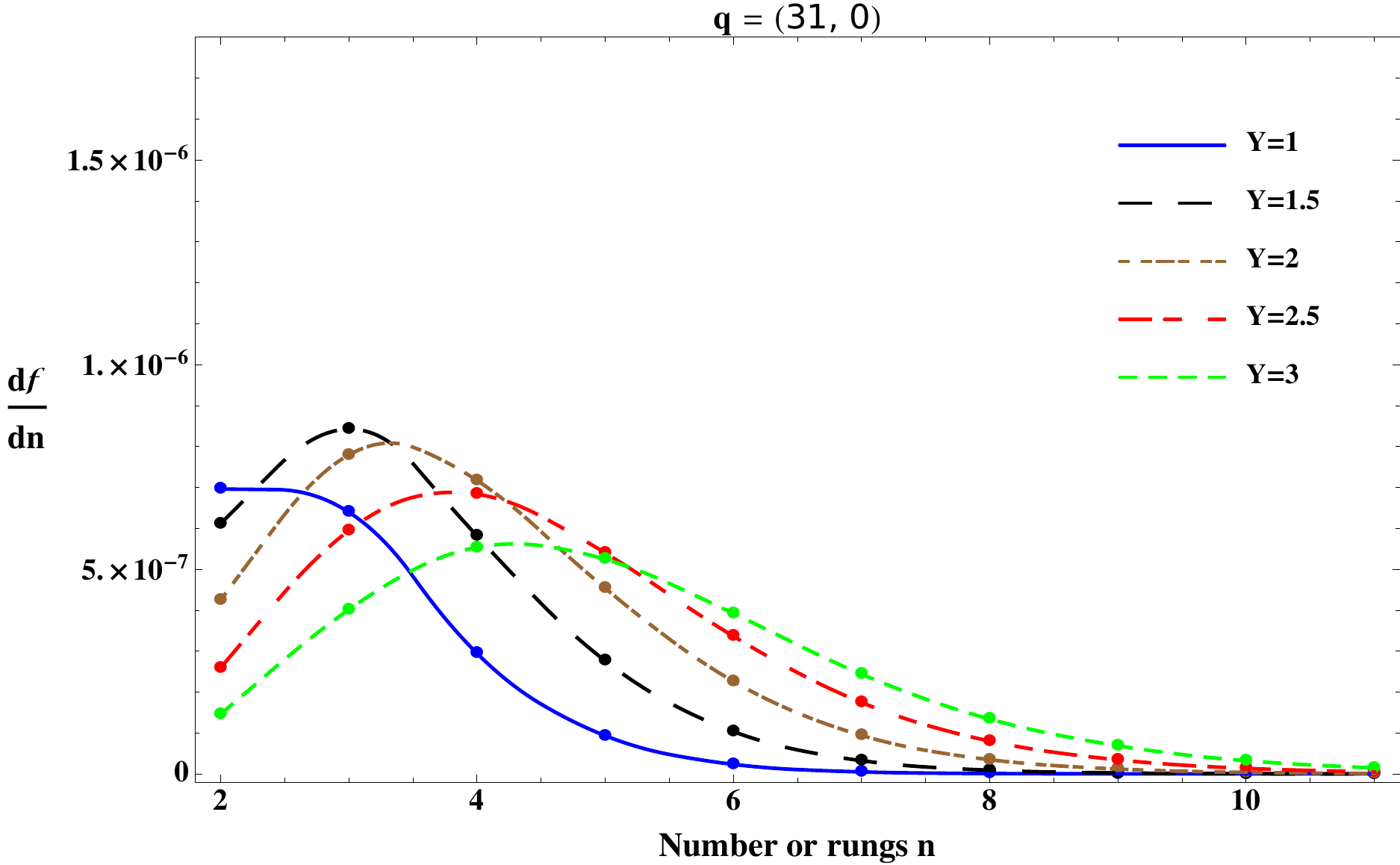}  \\
\vspace{.5cm}
    \includegraphics[width=0.75\linewidth]{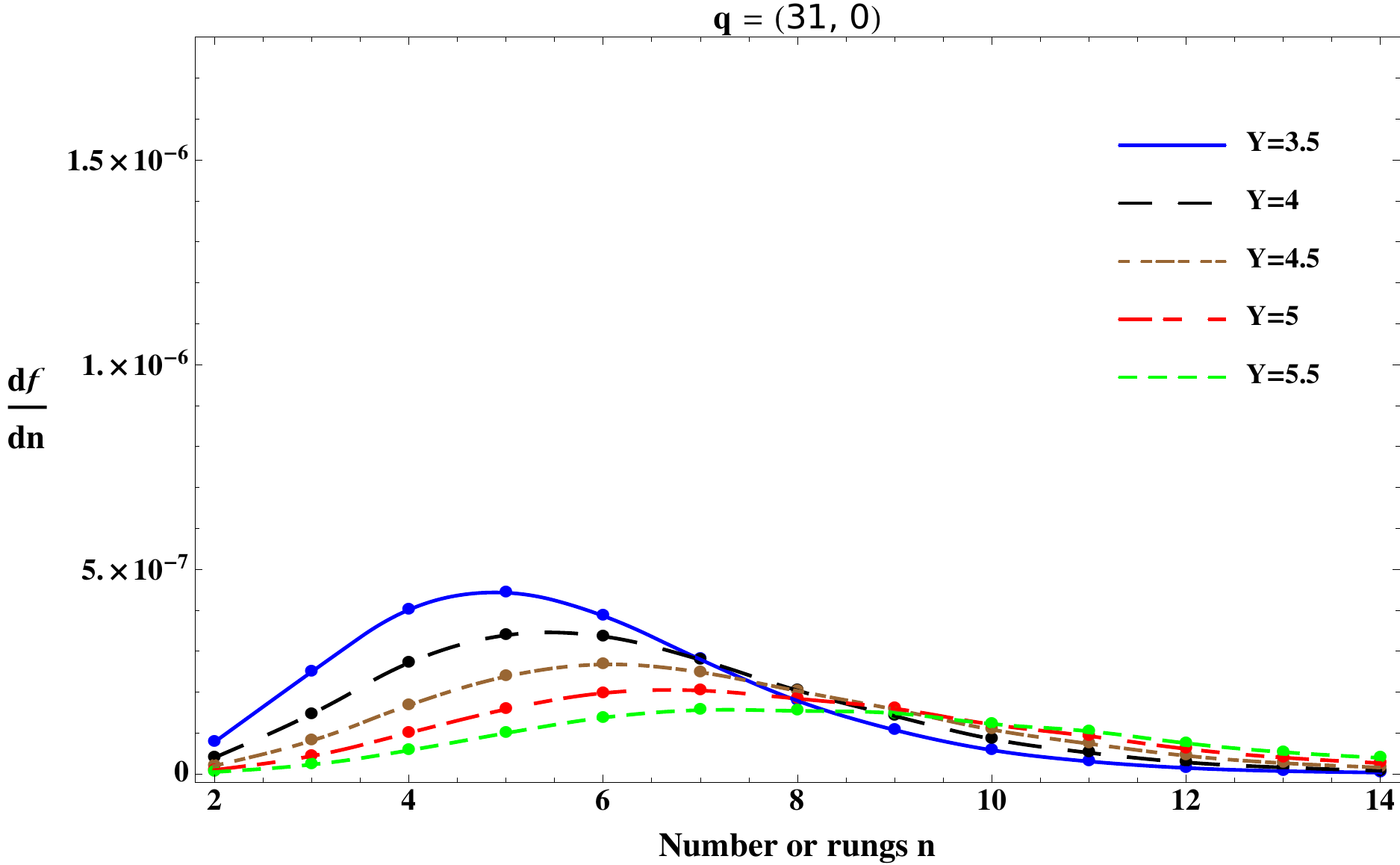}
\caption{{\it Multiplicity} plot for $q = 31$ GeV in the closed spin chain.}  
\label{q31-1}
\end{figure}

\begin{figure}
\centering
\includegraphics[width=0.75\linewidth]{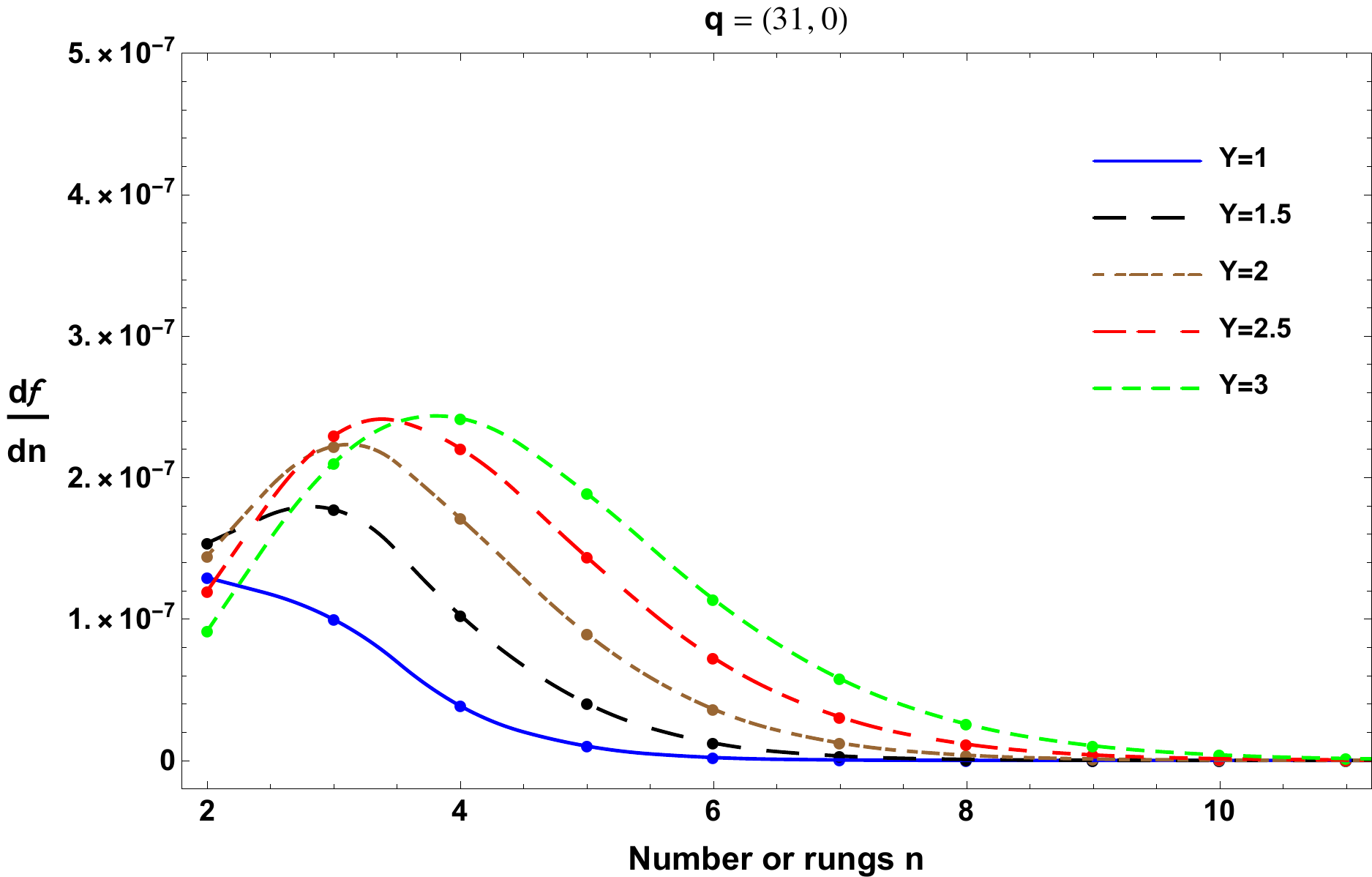}  \\
\vspace{.5cm}
    \includegraphics[width=0.75\linewidth]{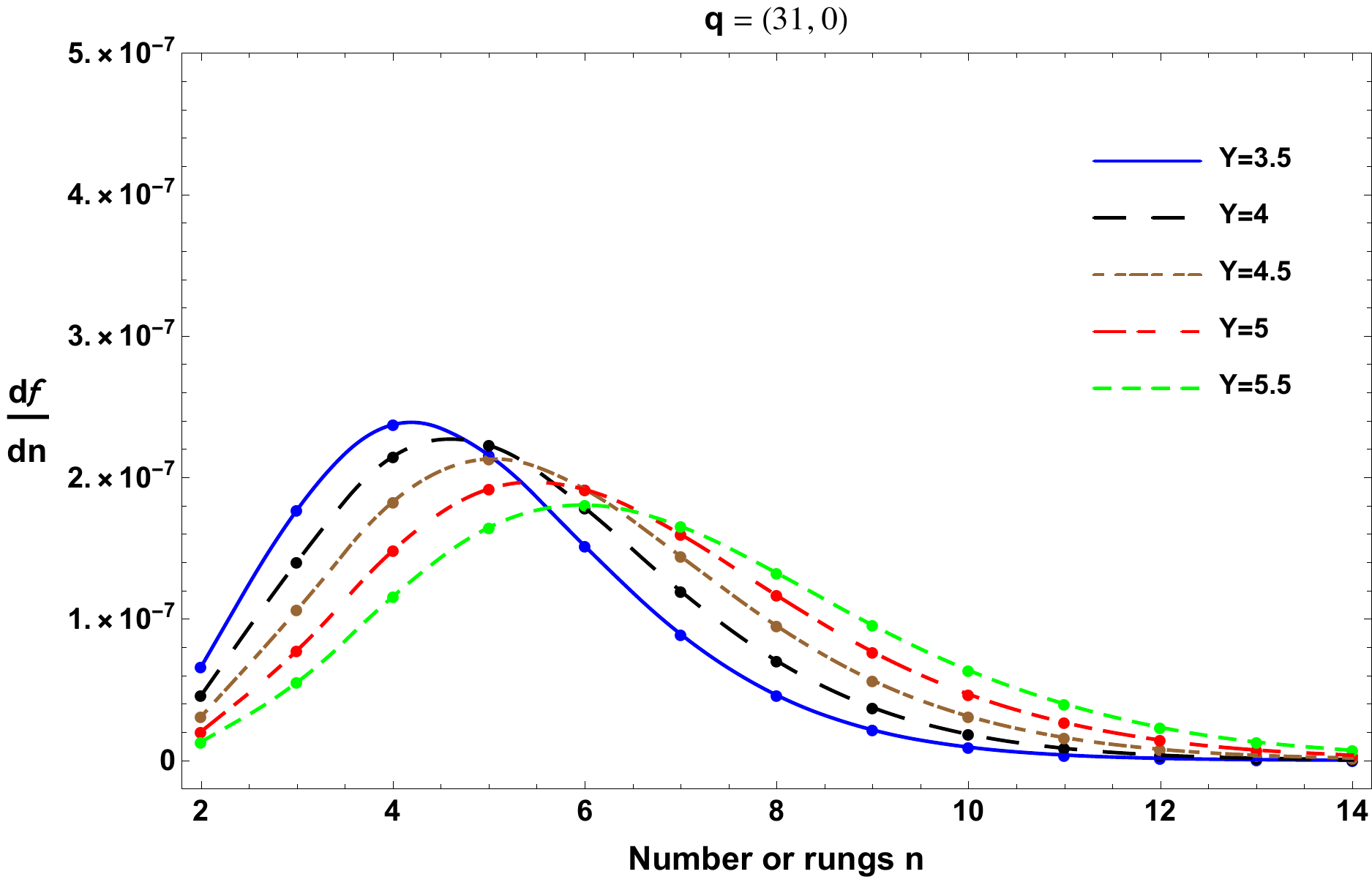}
\caption{{\it Multiplicity} plot for $q = 31$ GeV in the open spin chain.}  
\label{q31-2}
\end{figure}

In the following we are presenting four {\it multiplicity} (in the sense of counting the number of rungs in the diagram) plots, 
two of them (smaller and larger rapidities) with momentum transfer, ${\bf q} = (4,0)$ (Fig.~\ref{q4-2}) and two more for ${\bf q} = (31,0)$ (Fig.~\ref{q31-2}). We compare
these to the corresponding plots from the CSC, Figs.~\ref{q4-1} and~\ref{q31-1} respectively.

We verify anew that the contributions from each
iteration for a given $Y$ once plotted versus $n$, where $n$ is 
the number of rungs or iterations, follow a Poisson-like
distribution, similarly to the CSC. The similarities between the open and CSC do not end there. In both cases  we observe a global maximum at
some $n$ and then a decrease as $n$ increases. The Green function 
is noticeably smaller when the total momentum transfer is larger. Furthermore, the peak of the distribution moves similarly to larger values of $n$ for both values of ${\bf q}$ as 
$Y$ increases, whereas its height gets lower and the distributions
are much broader. As $|{\bf q}|$ increases the position of  the peak shifts to a larger $n$.

There is an interesting qualitative difference between the closed and OSC cases. Let us focus 
on the bottom plots in Figs.~\ref{q4-1} and~\ref{q4-2}. We observe that in the OSC  the decrease in the value of the maximal point of the distribution takes place very slowly as we vary $Y$. This is different to the CSC configuration where the distribution broadens very quickly with $Y$. This sudden broadening of the multiplicity distribution is a distinct signal of the cylinder topology of the contributing effective Feynman diagrams.  

\begin{figure}[H] 
    \centering
    \includegraphics[width=0.75\linewidth]{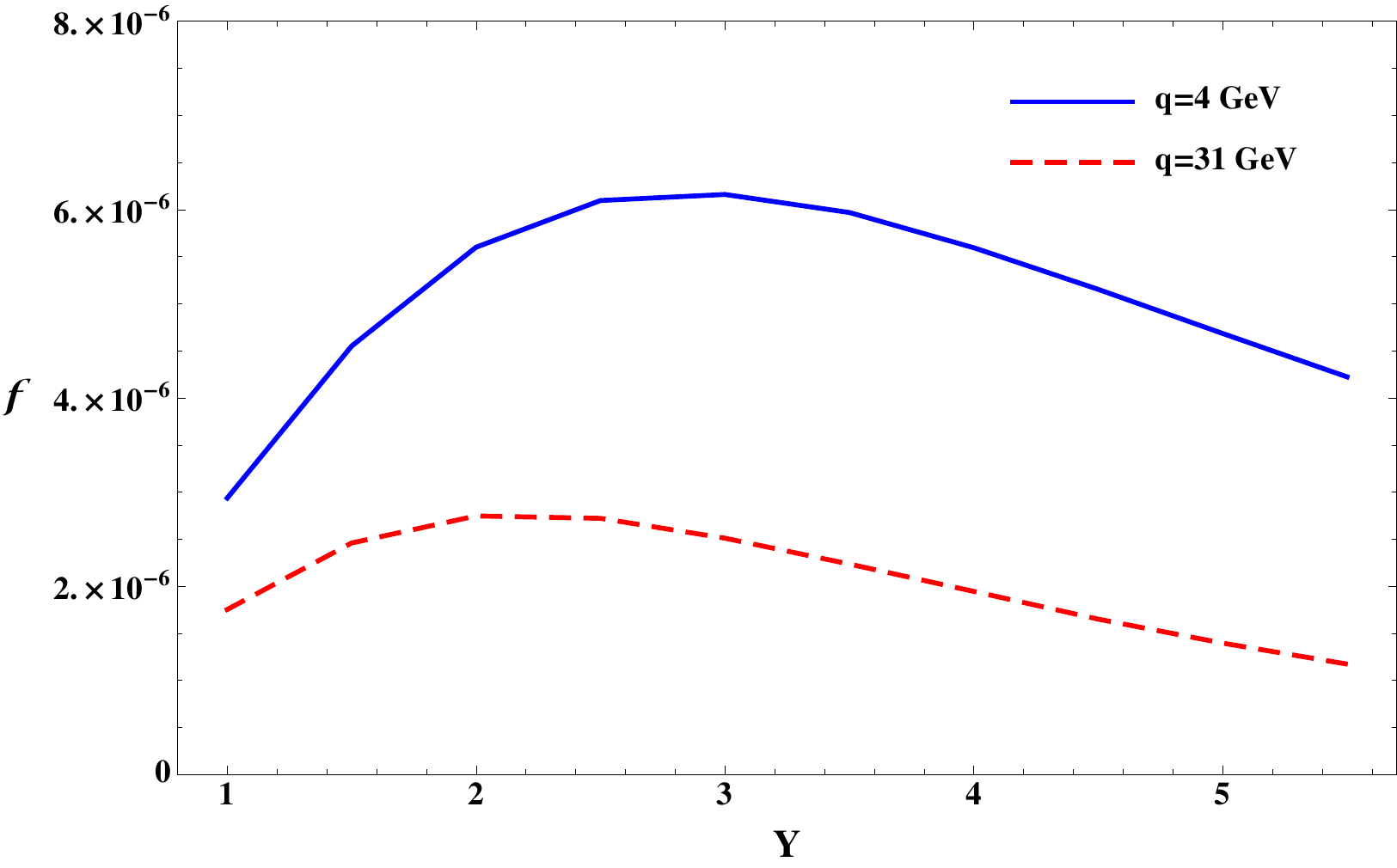} 
 \caption{Energy dependence of $f$ for $q = 4$ GeV (blue continuous line) and $q = 31$ GeV (red dashed line) for the closed spin chain.}
 \label{en1}
\end{figure}

The six-point gluon Green functions correspond to the areas under the distributions just described. Their dependence with $Y$ is drawn in Figs.~\ref{en1} and~\ref{en2}. While for the CSC we see that the energy ($\ln Y$) dependence  plot of the Green function has its peak at relatively small
rapidities ($Y<3$) for both ${\bf q} = (4, 0)$ and ${\bf q} = (31, 0)$ and then it starts to decrease noticeably fast,
for the OSC on the other hand, we see that for ${\bf q} = (4, 0)$ the curve increases monotonically.
If we now increase the momentum transfer, we see that for ${\bf q} = (17, 0)$ (brown dashed line)
the curve rises although much slower than for ${\bf q} = (4, 0)$. If we increase further
to ${\bf q} = (31, 0)$ (red dashed line) we see that for $Y>6$ (notice that we have pushed the upper limit of $Y$ here
for this plot to 6.5 units) the curve seems to decrease. An even further increase to ${\bf q} = (107, 0)$ (orange dashed line)
gives us a plot with a clear maximum at around $Y\sim3.4$.

To summarize, we find a similar qualitative behavior between the closed and open Green function for the main
features we could assess in our numerical analysis. However, the CSC Green function seems to approach
the asymptotia much faster than the OSC one. Moreover, the smaller the momentum transfer,
the more amplified this trend is. It will be very interesting to see what happens when one integrates the Green function with 
impact factors which is a crucial step to construct the full $n$-point amplitudes in the supersymmetric theory (Ref.~\cite{Bargheer:2016eyp} offers an interesting review on the systematics behind these calculations), but this is a question  beyond the scope of this work. Let us now conclude with a section devoted to a study of the graph complexity associated to the Feynman diagrams contributing to the gluon Green function. 

\section{Graph Complexity}

Let us highlight some aspects of graph theory which we have used in our study of the Reggeon spin chains (some reviews on the subject can be 
found in~\cite{GTWA,Gurau:2014vwa}). Our graphs ${\cal G}$ consist of a set of vertices $V$ and propagators (edges) $P$, ${\cal G}=(V,P)$. All the diagrams are connected since we do not allow for ${\bf p}_1$ to be equal to ${\bf p}_4$, ${\bf p}_2$ to ${\bf p}_5$, or ${\bf p}_3$ to ${\bf p}_6$. Only for connected graphs we can define {\it spanning trees}, which are those paths within the graph which connect all its vertices without any cycles. 

The {\it complexity} of an undirected connected graph corresponds to the number of all possible spanning trees of the graph ${\cal G}$. The {\it degree} of a particular vertex $i$, $d_{\cal G} (i)$, is the number of half-edges which are in contact with it. Since we can connect two or three propagators to each node then the vertices can have degree 2 or 3. 

If $|V|$ is the number of vertices in ${\cal G}$ then we can define the {\it degree matrix} of ${\cal G}$, $D_{\cal G}$, as the diagonal $|V| \times |V|$ square matrix with diagonal elements $d_{\cal G} (i)$. Its matrix elements are of the form $D_{\cal G} (i,j) = d_{\cal G} (i) \delta_{i j}$. As mentioned in a previous section, the  adjacency matrix of ${\cal G}$, $A_{\cal G} $, is the $|V| \times |V|$ square matrix with off-diagonal elements being the number of propagators connecting the vertex $i$ with the vertex $j$. The {\it Laplacian matrix} corresponds to the difference of these two matrices: $L_{\cal G} = D_{\cal G} - A_{\cal G}$. As we will see, these matrices carry the topological information of the graph. For the sake of clarity, we explicitly write the Laplacian matrices  associated to the diagrams in Fig.~\ref{grafs2}. In the CSC we have
\begin{eqnarray}
\left(
\begin{array}{cccccccccccc}
2&-1&0&0&-1&0&0&0&0&0&0&0\\
-1&2&-1&0&0&0&0&0&0&0&0&0\\
0&-1&3&-1&0&0&0&-1&0&0&0&0\\
0&0&-1&2&0&-1&0&0&0&0&0&0\\
-1&0&0&0&3&-1&-1&0&0&0&0&0\\
0&0&0&-1&-1&3&0&0&0&-1&0&0\\
0&0&0&0&-1&0&3&-1&0&0&-1&0\\
0&0&-1&0&0&0&-1&3&-1&0&0&0\\
0&0&0&0&0&0&0&-1&2&-1&0&0\\
0&0&0&0&0&-1&0&0&-1&3&0&-1\\
0&0&0&0&0&0&-1&0&0&0&2&-1\\
0&0&0&0&0&0&0&0&0&-1&-1&2\\
\end{array}
\right),
\end{eqnarray}
and in the OSC case, 
\begin{eqnarray}
\left(
\begin{array}{cccccccccccc}
2&-1&0&0&-1&0&0&0&0&0&0&0\\
-1&2&-1&0&0&0&0&0&0&0&0&0\\
0&-1&3&-1&0&-1&0&0&0&0&0&0\\
0&0&-1&2&0&0&0&-1&0&0&0&0\\
-1&0&0&0&3&-1&0&0&0&0&-1&0\\
0&0&-1&0&-1&3&-1&0&0&0&0&0\\
0&0&0&0&0&-1&3&-1&-1&0&0&0\\
0&0&0&-1&0&0&-1&3&0&-1&0&0\\
0&0&0&0&0&0&-1&0&3&-1&0&-1\\
0&0&0&0&0&0&0&-1&-1&2&0&0\\
0&0&0&0&-1&0&0&0&0&0&2&-1\\
0&0&0&0&0&0&0&0&-1&0&-1&2\\
\end{array}
\right).
\end{eqnarray}

\begin{figure}[ht] 
    \centering
    \includegraphics[width=0.8\linewidth]{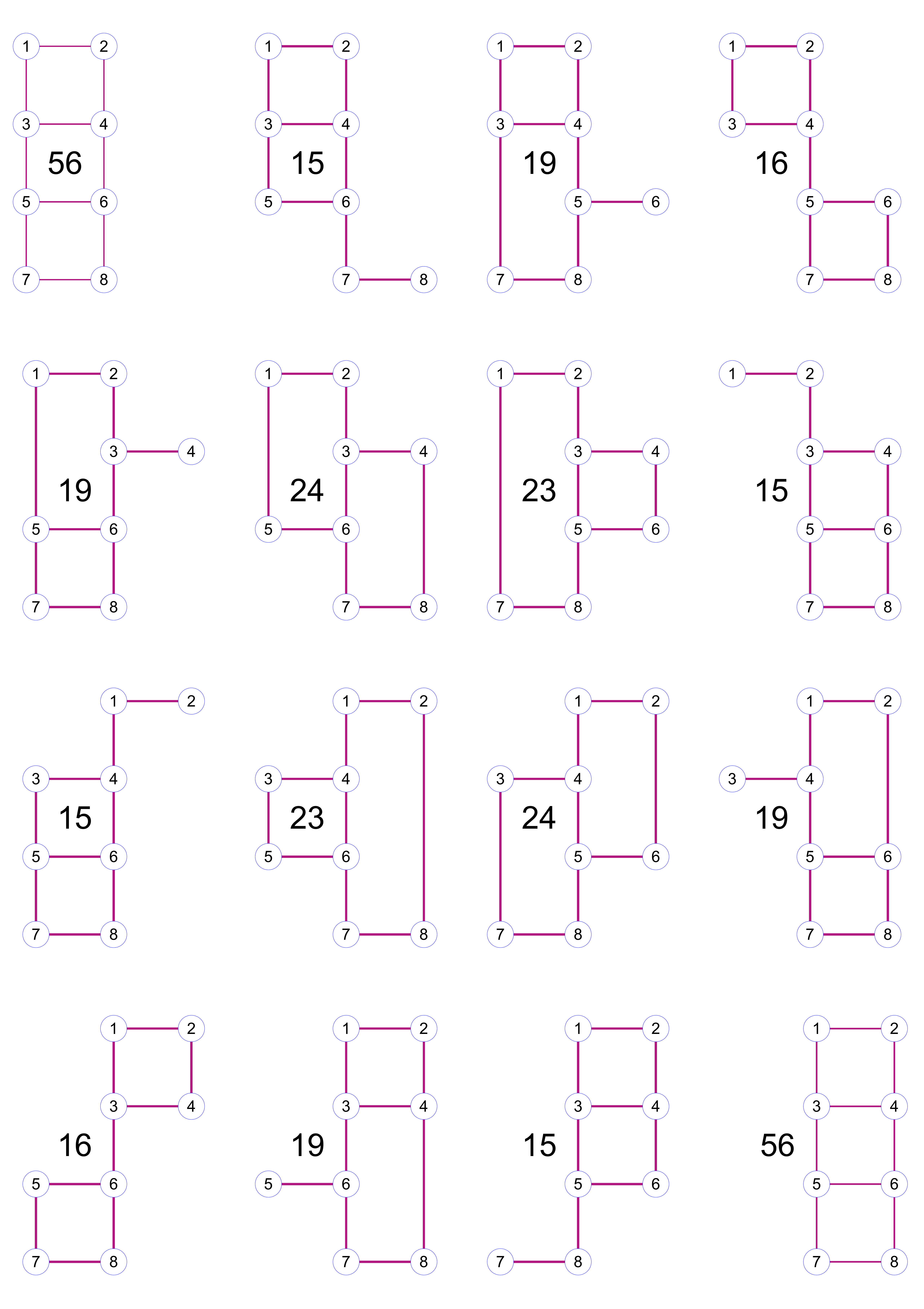} 
    \caption{Open spin chain topologies with 4 rungs. The labels of the topologies from left to right and top to bottom are
    {\small {
    \{L, L, L, L\}, \{L, L, L, R\}, \{L, L, R, L\}, \{L, L, R, R\}, \{L, R, L, 
  L\}, \{L, R, L, R\}, \{L, R, R, L\}, \{L, R, R, R\}, \{R, L, L, L\}, \{R, L, 
  L, R\}, \{R, L, R, L\}, \{R, L, R, R\}, \{R, R, L, L\}, \{R, R, L, R\}, \{R, 
  R, R, L\}, \{R, R, R, R\}}}.
   The corresponding complexity for each graph is shown.
    } 
    \label{4rungsdiagsopen} 
\end{figure}

\begin{figure}[ht] 
    \centering
    \vspace{-1cm}
    \includegraphics[width=0.9\linewidth]{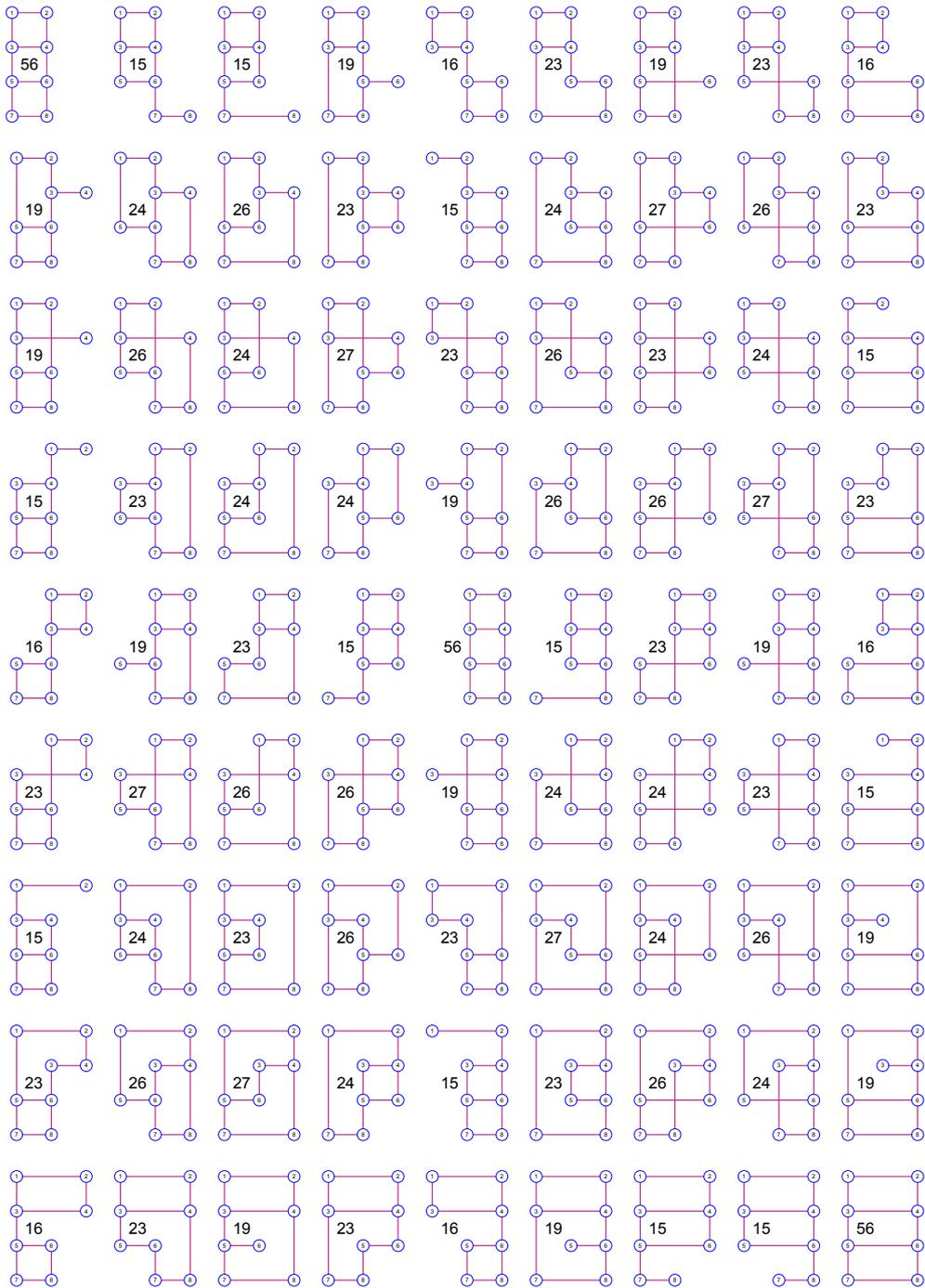} 
    \caption{Closed spin chain topologies with $n=4$ rungs. There are in total $3^n=81$ 
   topologies. Trivially, the set of topologies
    found in the OSC (Fig.~\ref{4rungsdiagsopen} ) is contained as a subset here.
    The corresponding complexity for each graph is shown.} 
    \label{4rungsdiagsclosed} 
\end{figure}

The Laplacian matrix is very useful to compute the complexity of each Reggeon graph in a  simple way. The key result is the well-known  {\it Matrix-Tree theorem} (Kirchhoff, 1847~\cite{Kirchhoff}) which is one of most fundamental results in combinatorial theory and states that the complexity of a graph corresponds to the value of the determinant of the Laplacian matrix once we remove one of its rows and one of its columns (the determinant of any of its principal minors). The complexity does not depend on a possible ordering of the vertices. Applying the Matrix-Tree theorem to the two graphs in  Fig.~\ref{grafs2} we find that the number of possible spanning trees in the CSC example is 532 and in the open case 463. For a fixed number of rungs the number of graphs in the CSC case is much larger than in the OSC configuration. As an example we show the relevant topologies along with their corresponding complexity for four rungs in Fig.~\ref{4rungsdiagsopen} for the OSC and in Fig.~\ref{4rungsdiagsclosed} for the OSC. It is natural to find that the OSC topologies are contained in the CSC possible diagrams. Moreover, the Pomeron ladder topology where all rungs are connecting the same pair of Reggeons appears twice in OSC (\{L,L,L,L\} and \{R,R,R,R\})
and thrice in CSC (\{L,L,L,L\}, \{M,M,M,M\} and \{R,R,R,R\}). It is interesting to note that the Pomeron ladder has always the maximal complexity $t(n)$ of all the 
diagram topologies for any given number $n$ of rungs. For example, in Figs.~\ref{4rungsdiagsopen}  and~\ref{4rungsdiagsclosed} $t(n) = 56$
for the Pomeron ladder topologies (see top left diagrams in both figures). The complexity of the Pomeron ladder is
equal to the number of spanning trees in  a $2 \times n$ grid which is given by
\begin{equation}
t(n) = 4 \, t(n-1) - t(n-2),  
\end{equation}
with t(0) = 0, t(1) = 1 or, equivalently, 
\begin{equation}
t(n) = \frac{\left(2+\sqrt{3}\right)^n-\left(2-\sqrt{3}\right)^n}{2 \sqrt{3}}\,.
\end{equation}
It will be relevant to investigate whether similar relations hold for the sub-leading in complexity
topologies in both the OSC and CSC cases.

\begin{figure}
\begin{center}
\begin{subfigure}{0.7\textwidth}
\includegraphics[height=6cm]{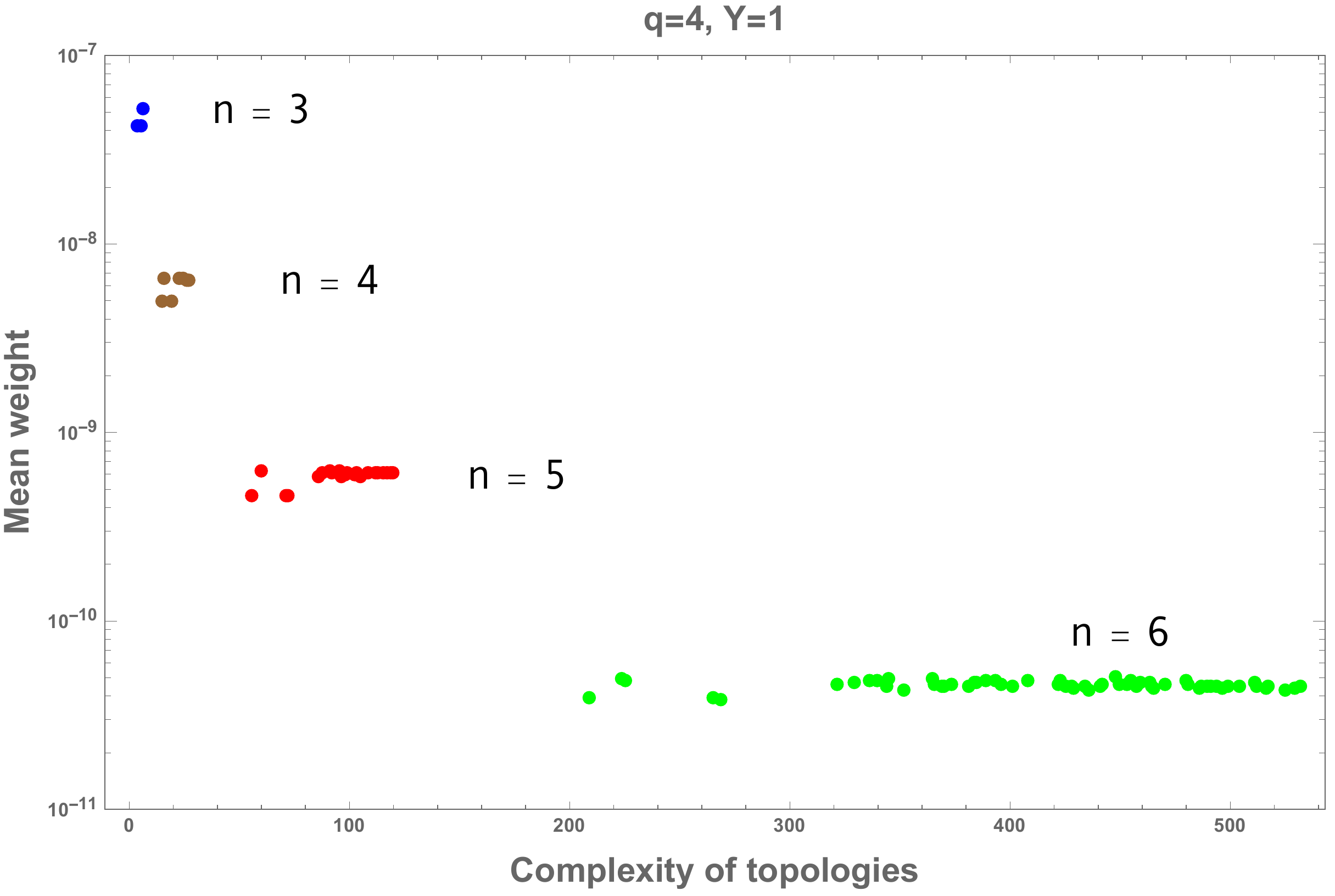}
\vspace{0.cm}
\caption{Closed spin chain, $Y=1$.} \label{ClosedY1}
\end{subfigure}
\begin{subfigure}{0.7\textwidth}
\includegraphics[height=6cm]{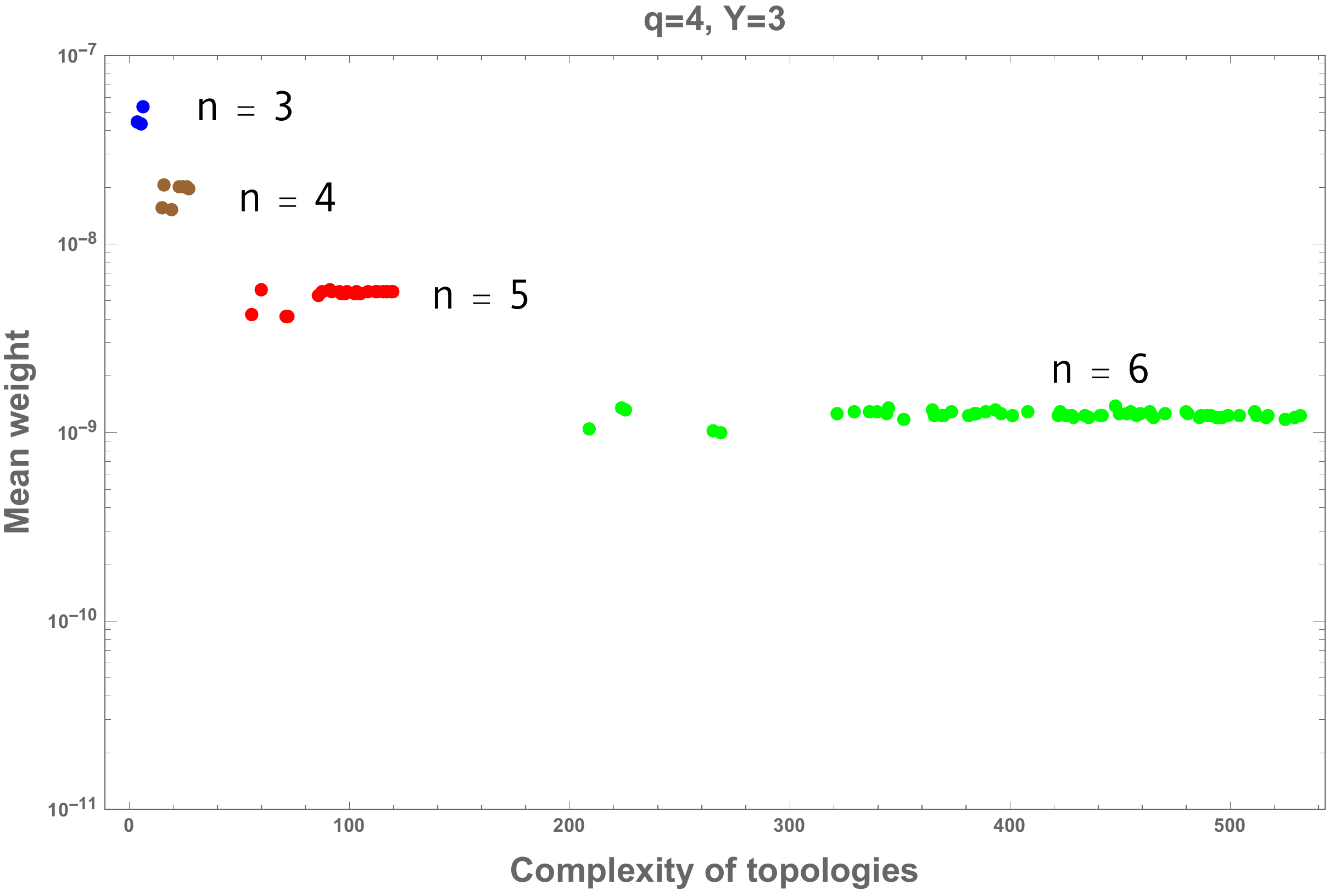}
\vspace{0.cm}
\caption{Closed spin chain, $Y=3$.} \label{ClosedY3}
\end{subfigure}
\begin{subfigure}{0.7\textwidth}
\includegraphics[height=6cm]{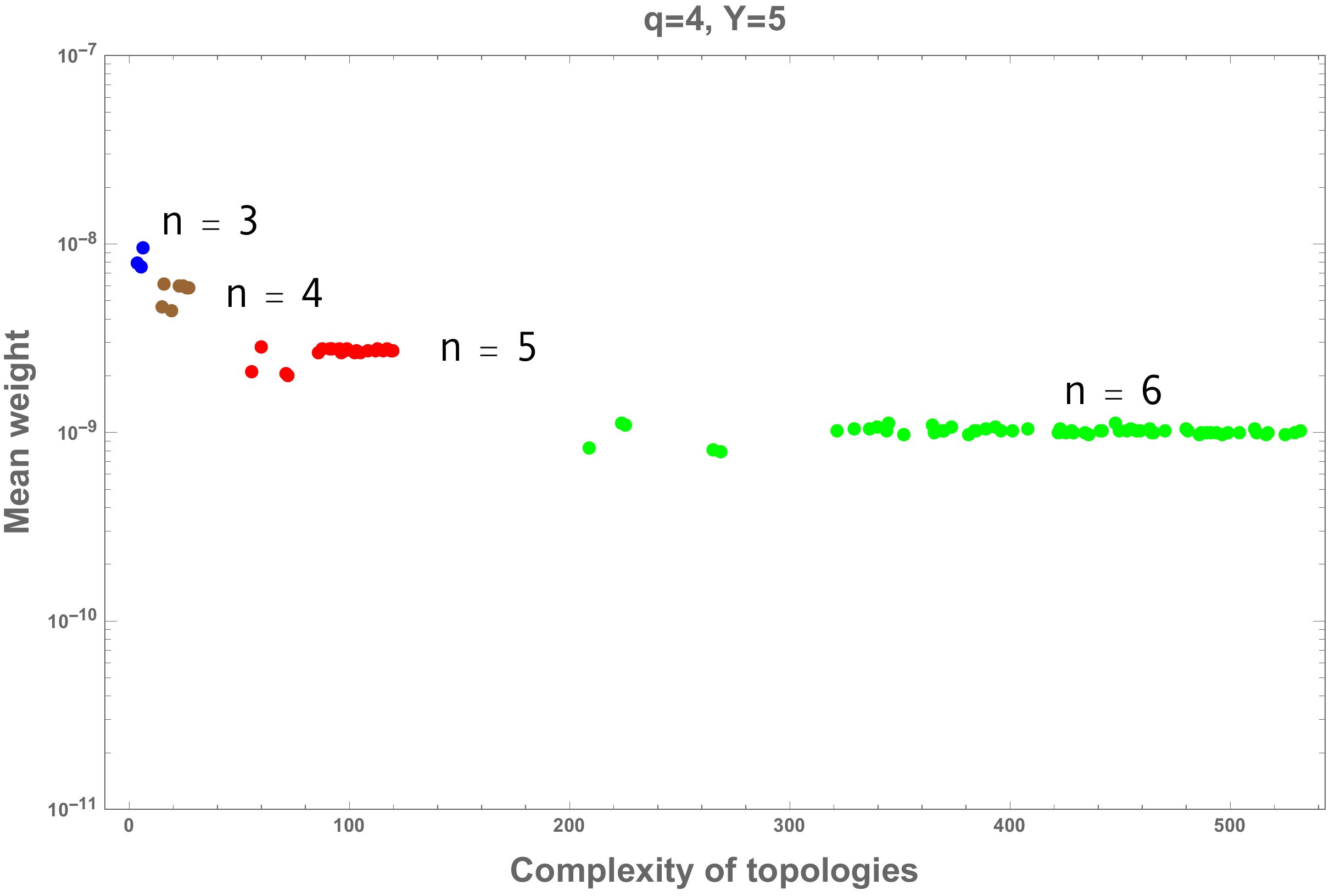}
\vspace{0.cm}
\caption{Closed spin chain, $Y=3$.} \label{ClosedY5}
\end{subfigure}
\end{center}
\caption{Average weight per complexity class for closed Reggeon webs.}
\label{ClosedComplexityPlots}
\end{figure}
\begin{figure}
\begin{center}
\vspace{-0.8cm}
\begin{subfigure}{0.7\textwidth}
\includegraphics[height=6cm]{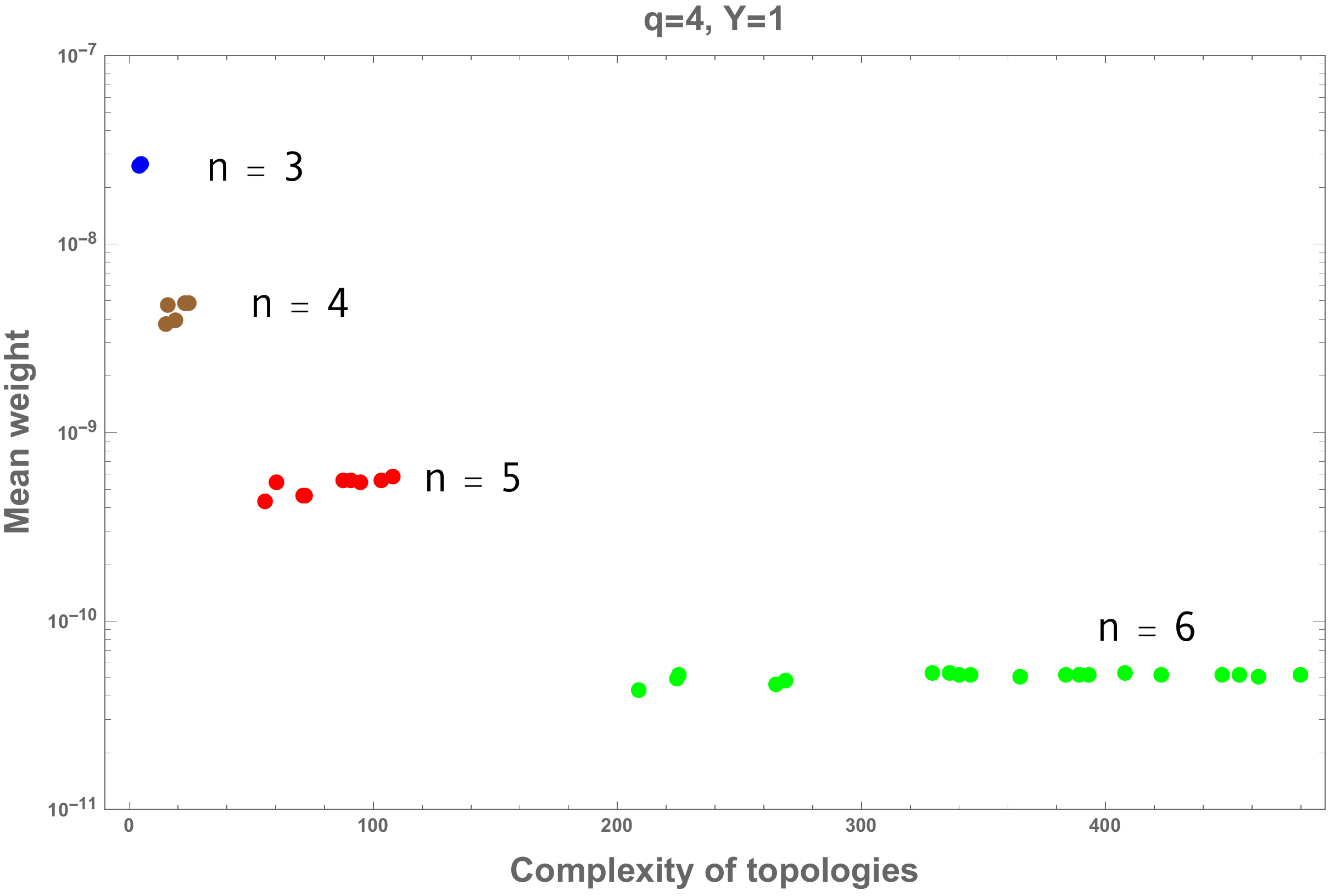}
\caption{Open spin chain, $Y=1$.} \label{ClosedY1}
\vspace{0.2cm}
\end{subfigure}
\begin{subfigure}{0.7\textwidth}
\includegraphics[height=6cm]{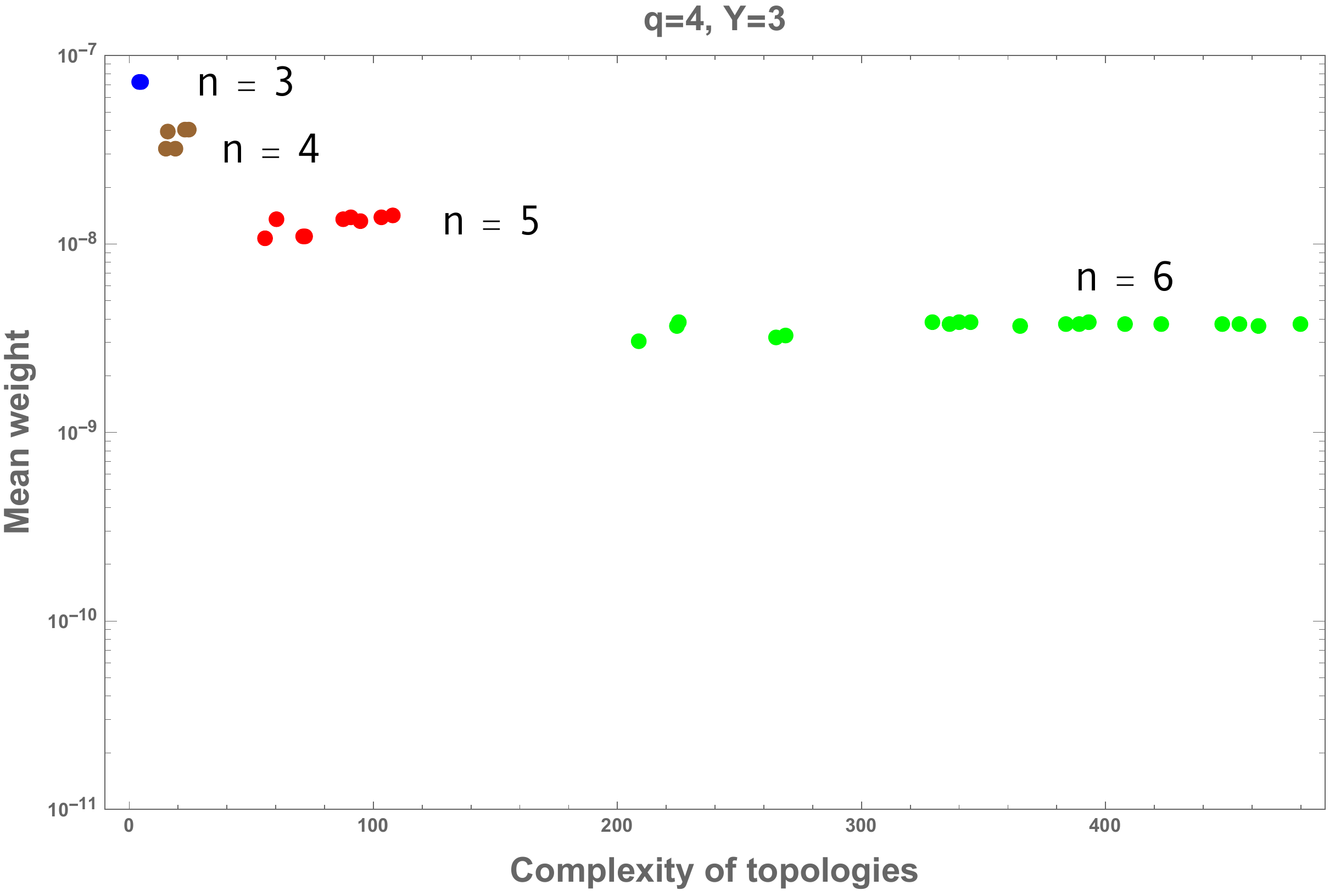}
\caption{Open spin chain, $Y=3$.} \label{ClosedY3}
\vspace{0.2cm}
\end{subfigure}
\begin{subfigure}{0.7\textwidth}
\includegraphics[height=6cm]{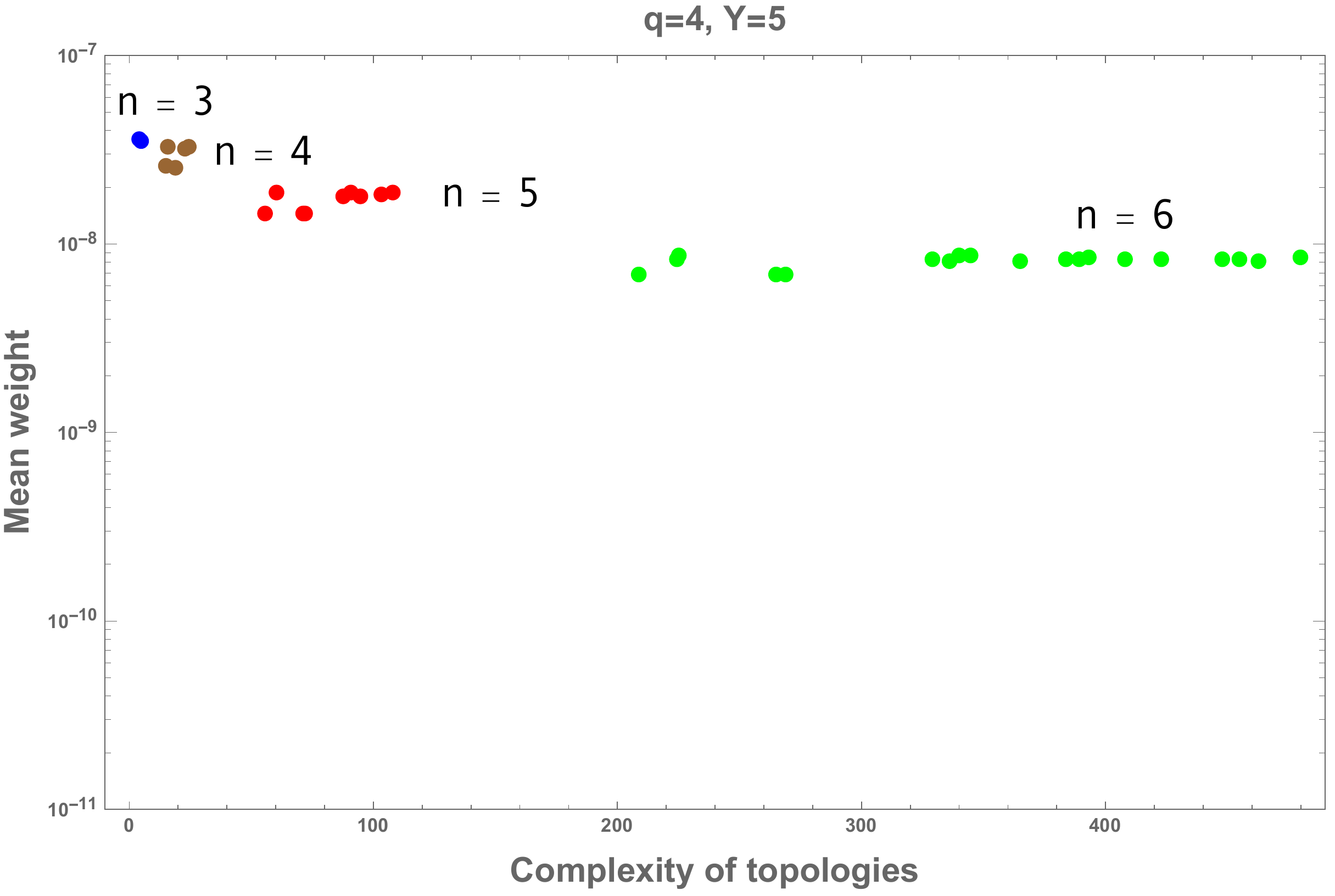}
\vspace{0.cm}
\caption{Open spin chain, $Y=5$.} \label{ClosedY5}
\end{subfigure}
\end{center}
\caption{Average weight per complexity class for open Reggeon webs.}
\label{OpenComplexityPlots}
\end{figure}

We have studied what is the contribution to the gluon Green function from different graph complexities. Of course each effective Feynman diagram carries a certain statistical weight in our Monte Carlo method to generate the solution to the BKP equation which is related to the particular number of reggeized gluon propagators and squared Lipatov's vertices present in a graph with $n$ rungs. It is clear that the average complexity of the graphs grows with $n$. We have found an interesting scaling behavior obtained by the following method. Let us consider all those diagrams with the same complexity for a fixed number of rungs. Then we evaluate the average weight of their contribution to the gluon Green function as a complexity class. All of this is calculated for a fixed value of the strong coupling and rapidity. We show a characteristic sample of our results in 
Fig.~\ref{ClosedComplexityPlots} for the CSC and in Fig.~\ref{OpenComplexityPlots} for the OSC. 

Per complexity value, those corresponding to a lower number of rungs have a bigger mean weight but their number is smaller. The amount of possible complexity values grows very fast with $n$. This is not surprising since the number of nodes in the graphs is proportional to the number of rungs. What we find very remarkable is that for the larger values of complexity in each set with the same $n$ all different complexities contribute with a very similar weight to the solution of the BKP equation. This trend is independent of having a closed or open Reggeon graph and for different values of the available parameters. We suspect that this ``complexity democracy" is likely to be related to the underlying integrability found by Lipatov. To establish the definite link is the subject of some of our current investigations.

\section{Conclusions and Outlook}

The high energy limit of scattering amplitudes in field theory and gravity~\cite{Bartels:2012ra} has a very rich structure. 
In QCD an integrable structure arises when the  center-of-mass energy is much larger than any Mandelstam invariant in the scattering process. This integrability, realized in coordinate representation, is related to the mapping to a Heisenberg ferromagnet with non-compact spins and periodic boundary conditions (closed spin chain). Recently, we have shown how to solve this system in momentum and rapidity space making use of Monte Carlo integration techniques. Our solution calculates the gluon Green function exactly and is general, it does not rely on any choice of normalization of the bound state wave function. For the three reggeon case it corresponds to the solution to the perturbative Odderon. 

More recently, Lipatov found a new integrable spin chain, in this case open, in the context of the calculation of scattering amplitudes in the $N=4$ supersymmetric Yang-Mills theory. This new open 
spin chain structure is important since it is the most complicated piece to evaluate when dealing with the multi-regge limit of scattering amplitudes in Mandelstam regions for amplitudes with a large number of legs. It is very important to understand it in detail in order to advance in our knowledge of the all-orders structure of amplitudes in a theory which allows for a smooth matching between the weak and the strong coupling limits~\cite{Bartels:2014mka}. 

In the present work we have solved the open spin chain problem exactly again using Monte Carlo integration. The infrared divergencies present in the calculation have been shown to factorize and we have investigated the infrared finite part of the gluon Green function. We have shown that it decreases with energy, as in the closed spin chain case although this behavior is delayed as the momentum transfer is reduced. Our results will allow to fix some of the uncertainties present when evaluating the 
eight-point amplitude in exact kinematics. For this it is still needed to integrate our results over impact factors and this will be the subject of our future work. Our techniques are valid for any number of reggeized gluons and apply also at next-to-leading and higher orders~\cite{Bartels:2012sw}. This implies that they will be important for the evaluation of the general $n$-point amplitudes in exact kinematics~\cite{Dixon:2014voa}. 

As a by-product of our work, we have found an intriguing scaling behavior of what we can call 
weighted complexity of the Feynman graphs contributing to the gluon Green function. The complexity of a diagram is a well-defined quantity in graph theory. We have evaluated the average weight per topology in the sense of its total contribution to the gluon Green function and found that it is approximately constant for a fixed number of rungs of the class of effective Feynman diagrams. This ``complexity democracy" is very likely related to the integrability found by Lipatov. It will be interesting to find the precise link between both concepts.

\vspace{1cm}
\begin{flushleft}
{\bf \large Acknowledgements}
\end{flushleft}
We acknowledge support from the Spanish Government grants FPA2015-65480-P, FPA2016-78022-P and Spanish MINECO Centro de Excelencia Severo Ochoa Programme (SEV-2016-0597).

\end{document}